\documentclass[paper]{JHEP3}
\JHEPspecialurl{http://jhep.sissa.it/JOURNAL/JHEP3.tar.gz}

\usepackage{amsmath}
\usepackage{amssymb}
\usepackage{bbm}
\usepackage{epsfig}
\usepackage{multicol}
\usepackage{dcolumn}
\usepackage{booktabs}
\usepackage{here}

\newcommand {\Tr}{\mathop{\rm Tr}}

\newcommand {\dd}{\mbox{d}}

\newcommand {\SO}{\mathop{\rm SO}}
\newcommand {\SU}{\mathop{\rm SU}}

\newcommand{\beq}{\begin{equation}}
\newcommand{\eeq}{\end{equation}}
\newcommand{\beqa}{\begin{eqnarray}}
\newcommand{\eeqa}{\end{eqnarray}}

\newcommand{\tri}{\scalebox{1}[0.8]{$\vartriangle$}}
\newcommand{\btri}{\scalebox{1}[0.8]{$\blacktriangle$}}
\newcommand{\dtri}{\scalebox{1}[0.8]{$\triangledown$}}
\newcommand{\dbtri}{\scalebox{1}[0.8]{$\blacktriangledown$}}
\newcommand{\sq}{\scalebox{1}[0.6]{$\lozenge$}}
\newcommand{\bsq}{\scalebox{1}[0.6]{$\blacklozenge$}}

\def\theenumi{\roman{enumi}}

\title{%
Systematic study of the SO(10) symmetry breaking vacua
in the matrix model for type IIB superstrings
}

\author{Jun Nishimura \\
  KEK Theory Center, 
  Institute of Particle and Nuclear Studies, \\
  High Energy Accelerator Research Organization,
  and \\
  Department of Particle and Nuclear Physics, \\
  The Graduate University for Advanced Studies (SOKENDAI), \\
  1-1 Oho, Tsukuba 305-0801, Japan \\
  E-mail: \email{jnishi@post.kek.jp}
}
\author{Toshiyuki Okubo \\
  Faculty of Science and Technology,
  Meijo University, \\
  Nagoya, 468-8502, Japan \\
  E-mail: \email{tokubo@meijo-u.ac.jp}
}

\author{Fumihiko Sugino \\
Okayama Institute for Quantum Physics\\
Kyoyama 1-9-1, Kita-ku, Okayama 700-0015, Japan\\
  E-mail: \email{fumihiko\_sugino@pref.okayama.lg.jp}
}

\preprint{%
KEK-TH-1483\\OIQP-11-06
}

\abstract{
We study the properties of the space-time that 
emerges \emph{dynamically} from
the matrix model for type IIB superstrings in ten dimensions.
We calculate the free energy and the extent of space-time
using the Gaussian expansion method up to the third order.
Unlike previous works, we study the
SO($d$) symmetric vacua
with all possible values of $d$
within the range $2 \le d \le 7$,
and observe clear indication of 
plateaus in the parameter space of the Gaussian action,
which is crucial for the results to be reliable.
The obtained results indeed exhibit
systematic dependence on $d$,
which turns out to be surprisingly similar to 
what was observed recently in an analogous work on the
six-dimensional version of the model.
In particular, we find the following properties: 
i) the extent in the shrunken directions is given
by a constant, which does not depend on $d$;
ii) the ten-dimensional volume of 
the Euclidean space-time is given
by a constant, which does not depend on $d$ except for $d=2$;
iii) The free energy takes the minimum value at $d=3$.
Intuitive understanding of these results is given
by using the low-energy effective theory and some Monte Carlo results.
}

\keywords{Matrix Models, Superstring Vacua}

\begin{document}

\section{Introduction}
\label{sec:intro}

In recent years a lot of efforts have been devoted to constructing
a stable (or 
long-lived meta-stable) vacuum
in string theory, which is appropriate to describe
our real world. In this kind of approach, however,
there is no objective criterion to pick up one of the
tremendously many possible vacua,
which
is commonly referred to as the landscape problem 
in the literature.
%
It would be
of course 
scientifically more desirable if
a unique vacuum is chosen nonperturbatively and 
the chosen vacuum actually describes our real world.
The aim of the present paper is to show that this possibility
should not be totally forgotten as theorists'\ dream.
Our explicit calculations rather suggest that we might be quite
close to it although a big twist may still be needed to achieve
the final goal.

In order to address such an issue,
we certainly need 
a nonperturbative formulation of superstring theory.
There are actually quite a few proposals made
in the late 90s \cite{Banks:1996vh,Ishibashi:1996xs,Dijkgraaf:1997vv}
after the discovery of D-branes.
In this paper we study 
the type IIB matrix model \cite{Ishibashi:1996xs},
which is proposed as a nonperturbative formulation of 
type IIB superstring theory in ten dimensions.
This model can be obtained formally by taking
the zero-volume limit of $\SU(N)$ super Yang-Mills theory 
in ten dimensions.\footnote{This relationship
to $\SU(N)$ super Yang-Mills theory is considered at the level of
classical action. Therefore the well-known gauge anomaly
in the ten-dimensional super Yang-Mills theory is not an issue.
} 
The integer $N$ can be viewed as
a sort of regularization parameter, which should be taken to infinity
eventually.
In this formulation of superstring theory, 
the ten-dimensional target space
is represented by
the ten bosonic Hermitian matrices \cite{Aoki:1998vn}, 
which originate from the ten-dimensional gauge field
in the super Yang-Mills theory. 
The model has manifest SO(10) invariance,\footnote{This should be
considered as a virtue of this model, which is important for
the issue we address in this paper. In any other proposal
for a nonperturbative formulation of superstring/M theory, 
the space-time symmetry is not manifestly preserved.} 
which is inherited from the super Yang-Mills theory.
The question we would like to ask is whether
this SO(10) symmetry is spontaneously broken in the large-$N$ limit
down to SO(4), which is the symmetry corresponding to 
our space-time.\footnote{See ref.~\cite{ncym} for discussions 
on the idea of the emergent gravity,
which is deeply related to the above scenario.}

In ref.\ \cite{Nishimura:2001sx} two of the authors
applied the Gaussian expansion method to this issue.
The free energy of the SO($d$) symmetric vacua
for $d=2,4,6,7$
was calculated up to the third order of the expansion,
and $d=4$ was found to give the smallest value.
Moreover, the extent of space-time
in the $d$ directions was found to be
larger than those in the remaining $(10-d)$ directions.
This result motivated higher order 
calculations up to the eighth order 
\cite{Kawai:2002jk,Kawai:2002ub,Aoyama:2006di,Aoyama:2006rk,Aoyama:2006je}
for the $d=4$ and $d=7$ cases.
While the results revealed interesting qualitative differences
between the two cases, 
the situation turned out to be obscure.

In the Gaussian expansion method,
the result depends on free parameters
introduced in the Gaussian action.
The crucial point is that one can still
make a reliable prediction
by identifying the ``plateau region''
in the parameter space,
in which the result becomes almost constant.
A practical approach
is to obtain the points in the parameter space
at which the result becomes stationary.
As one goes to higher orders, one typically
obtains more and more stationary points giving totally different results.
However, if it turns out that 
there are quite a few points that give
approximately the same results, one may regard it as
indication of the plateau region.
Such behaviors were indeed observed in various simple models
and the obtained results confirmed the validity of the method. 
(For example, see 
refs. \cite{Kawai:2002jk,Nishimura:2002va,Nishimura:2003gz}.) 
However, in the case of type IIB matrix model, 
the plateau region has not yet been identified unambiguously.
%
%

Recently it was suggested \cite{Aoyama:2010ry}
that the above situation might be
due to the extra symmetry $\Sigma_d$,
which was imposed on the shrunken $(10-d)$ directions
in order to make the calculations feasible.
By imposing a symmetry 
${\rm SO}(d) \times \Sigma_d \subset {\rm SO}(10)$,
which is stronger than just ${\rm SO}(d)$,
one can reduce the number of free parameters considerably.
While this makes it much easier to obtain
stationary points,
it also makes
the available stationary points strongly restricted
by the chosen extra symmetry $\Sigma_d$.
As a result, one might miss the opportunity to observe 
clear indication of plateaus for the SO($d$) symmetric vacua
that might otherwise be there.

%

This possibility was noticed in a similar study for
the six-dimensional version of 
the type IIB matrix model \cite{Aoyama:2010ry}.
The model can be obtained formally by
the zero-volume limit of $\SU(N)$ super Yang-Mills theory 
in six dimensions, and it is expected to have properties
analogous to the type IIB matrix model
such as the spontaneous breakdown of the SO(6) symmetry.
On the technical side, 
the analysis becomes much easier than in the type IIB matrix model,
and it was possible to perform calculations 
for all the values of $d$ within the range $2 \le d \le 5$.
Moreover, calculations were done
imposing \emph{only the SO($d$) symmetry}
up to order 3 for $d=3,4,5$, and 
up to order 5 for $d=4,5$.
Quite a few stationary points were found for each $d$,
and clear indication of plateaus was observed.
This enabled reliable predictions for both the free energy
and the extent of space-time,
which
exhibited interesting qualitative features summarized
as follows.
\begin{enumerate}
\item The extent of space-time in the shrunken directions
is given by a
constant ($r$), which is independent of $d$. 
(universal ``compactification'' scale)
\item The six-dimensional volume of the Euclidean space-time is 
given by a
constant ($v\equiv\ell^6$),
which is independent of $d$ except for $d=2$. (constant volume property)
\item The free energy takes the minimum value at $d=3$. \\
\end{enumerate}
Intuitive understanding for these properties is also given
in ref.~\cite{Aoyama:2010ry}.
The properties i) and ii) can be understood 
by considering the low-energy effective theory, which is given 
in terms of a system similar to the 
branched polymer \cite{Aoki:1998vn,Ambjorn:2000dx}.
The two dynamical scales $r$ and $\ell$, 
which characterize the properties i) and ii), respectively,
are reproduced numerically by Monte Carlo simulation \cite{AAN}.
The property iii) as well as the anomaly for $d=2$ in the property ii)
can be
understood from the properties
of the fermion determinant \cite{NV}.

With all these new insights,
we redo the calculations for
the type IIB matrix model
and study the SO($d$) symmetric vacua for all values of $d$
within the range $2\le d \le 7$.
Unlike in the six-dimensional case,
it is difficult to perform calculations
imposing \emph{only the SO($d$) symmetry}.
However, 
the calculations in
the six-dimensional case \cite{Aoyama:2010ry} 
suggest
that the stationary points 
obtained in the plateau region
have all the variety of symmetries in the shrunken directions.
Therefore, we perform calculations in the ten-dimensional case
imposing ${\rm SO}(d) \times \Sigma_d \subset {\rm SO}(10)$
with various possible $\Sigma_d$ for each $d$.
In fact we exhaust all the possible extra symmetries that leave
not more than five free parameters in the Gaussian action.
By combining all the solutions obtained in this way,
we were able to observe clear indication
of plateaus already at the 3rd order calculations
as we did in the six-dimensional case.
This should be contrasted to the situation with the previous 
calculations for the type IIB matrix model, 
where one particular symmetry $\Sigma_d$ 
was chosen for each $d$, and hence the number of solutions was not 
enough to clearly identify the plateau.

The free energy and the extent of space-time
obtained for each $d$ in the plateau region
indeed exhibit systematic $d$ dependence
analogous to i)-iii) observed for the six-dimensional case.
This is understandable theoretically
given that the low-energy
effective theory is described by a similar
branched-polymer-like system \cite{Aoki:1998vn,Ambjorn:2000dx}
and that the fermion determinant has similar properties \cite{NV}.
Let us emphasize, however, that the Gaussian expansion method
knows neither of these facts, and yet it revealed the same qualitative
behaviors of the two models. 
Moreover, the free energy
obtained for the SO($d$) symmetric vacua 
is actually quite 
close to the value obtained from the 
formula for the partition function conjectured by
Krauth, Nicolai and Staudacher \cite{Krauth:1998xh},
\emph{for both the 6d and 10d models}.
We consider these as strong evidence
for the validity of the present calculations.

The rest of this article is organized as follows. 
In section \ref{sec:model} we define the model and
the observable which serves as an order parameter
for the spontaneous breaking of rotational SO(10) symmetry.
In section \ref{sec:gaussian} we describe the method we 
use to study the model.
In section \ref{sec:result}
we present our results for the free energy
and the extent of space-time 
in the SO($d$) symmetric vacua ($ 2 \le d \le 7$).
Section \ref{sec:discussion} is devoted to a summary and
discussions.
In appendix \ref{sec:app_ansatz} we present the details of the
ansatz we use to study each of the ${\rm SO}(d)$ symmetric vacua.
In appendix \ref{sec:KNSresult} we derive the value of 
free energy from the formula for the partition function
conjectured by Krauth, Nicolai and Staudacher.

\section{The model and the order parameter}
\label{sec:model}

%
%

 
%
%
%

The type IIB matrix model can be obtained formally
by the zero-volume limit of 
$D=10$ 
$\SU(N)$ pure super Yang-Mills theory,
and its partition function is given by 
\begin{eqnarray}
  Z &=& \int \dd A \, \dd \Psi
\, e^{-S_{\rm b} - S_{\rm f}} \ , 
\label{eq:10dpf} \\
  S_{\rm b} &=& - \frac{1}{4 g^2}  \Tr [ A_\mu, A_\nu ]^2  \ , 
\label{eq:sb} \\
  S_{\rm f} &=& - \frac{1}{2 g^2} 
\Tr\left(\Psi_\alpha (C \Gamma^\mu)_{\alpha\beta}
    [ A_\mu, \Psi_\beta ] \right) \ . 
\label{eq:sf}
\end{eqnarray}
Here $A_\mu$ ($\mu = 1,\cdots,10$) are 
traceless $N\times N$ Hermitian matrices, 
whereas $\Psi_\alpha$ ($\alpha = 1,\cdots,16$) are 
traceless $N\times N$ matrices with Grassmannian entries. 
The parameter $g$ can be scaled out by appropriate redefinition
of the matrices, and hence it is just a scale parameter
rather than a coupling constant.
We therefore
set $g^2 N = 1$ from now on
without loss of generality.
The integration measure for $A_\mu$ and $\Psi_\alpha$
is given by
\beq
  \dd A 
=
  \prod_{a=1}^{N^2-1} \prod_{\mu=1}^{10} \frac{\dd A_\mu^a}{\sqrt{2\pi}} \ , 
\quad \quad
  \dd \Psi
=
 \prod_{a=1}^{N^2-1} \prod_{\alpha=1}^{16} 
    \dd \Psi_\alpha^a \ ,
\label{eq:measure_psi}
\eeq
where $A_\mu^a$ and $\Psi_\alpha^a$ are 
the coefficients in the expansion 
$A_\mu = \sum_{a=1}^{N^2-1} A_\mu^a T^a$ \emph{etc.}
with respect to the $\SU(N)$ generators $T^a$ 
normalized as $\Tr (T^a T^b) = \frac{1}{2}\delta^{ab}$. 

The model has an $\SO(10)$ symmetry, 
under which $A_\mu$ and $\Psi_\alpha$ transform as a vector
and a Majorana-Weyl spinor, respectively.
The $16\times 16$ matrices $\Gamma_\mu$ are the gamma matrices after 
the Weyl projection, and $C$ is the charge conjugation matrix,
which satisfies $(\Gamma_\mu)^T = C\Gamma_\mu C^\dag$ and $C^T = C$.


In order to discuss the spontaneous symmetry breaking (SSB)
of $\SO(10)$ in the large-$N$ limit, 
we consider the 
``moment of inertia'' tensor \cite{Aoki:1998vn,Hotta:1998en}
\begin{equation}
  T_{\mu\nu} = \frac{1}{N} \Tr (A_\mu A_\nu) \ , 
\label{eq:tmunu}
\end{equation}
which is a $10\times10$ real symmetric tensor. 
We denote its eigenvalues as $\lambda_j$ ($j=1,\cdots, 10$) 
with the specific order 
\begin{equation}
  \lambda_1 \geq \lambda_2 \geq \cdots \geq \lambda_{10}  \ .
\label{eq:lambda}
\end{equation}
If the SO(10) is not spontaneously broken, 
the expectation values $\langle \lambda_j  \rangle$ 
($j=1, \cdots , 10$) should be all equal in the large-$N$ limit.
Therefore, if we find that they are not equal,
it implies that the SO(10) symmetry is spontaneously broken.
Thus the expectation values $\langle \lambda_j  \rangle$ 
serve as an order parameter of the SSB.
{}In ref.\ \cite{NV} it was found that 
the phase of the fermion determinant (or Pfaffian, strictly speaking)
favors $d(\ge 3)$-dimensional configurations,
which have
$ \lambda_j  $ ($j=d+1, \cdots , 10$) much
smaller than the others.
This suggests the possibility that
the $\SO(10)$ symmetry is broken down to $\SO(d)$ with $d \ge 3$.
Since the eigenvalue distribution of $A_\mu$ represents 
the extent of space-time in the type IIB matrix model \cite{Aoki:1998vn},
the above situation realizes
the dynamical compactification to $d$-dimensional space-time.

In general one can obtain
supersymmetric matrix models
by taking the zero-volume limit of 
pure super Yang-Mills theories
in $D=3$, 4, 6 and 10 dimensions,
where the $D=10$ case corresponds to the type IIB matrix model.
The convergence of the partition function for general $D$
was investigated
both numerically \cite{Krauth:1998xh} and analytically \cite{AW}. 
The $D=3$ model is ill-defined since the partition function
is divergent. The $D=4$ model has a real positive
fermion determinant, and Monte Carlo simulation
suggested the absence of the SSB of rotational symmetry \cite{Ambjorn:2000bf}.
(See also refs.\ \cite{Burda:2000mn,Ambjorn:2001xs}.) 
The $D=6$ model and the $D=10$ model both
have a complex fermion determinant,
whose phase is expected to play a crucial 
role \cite{NV,sign,Nishimura:2001sq,Nishimura:2004ts,Anagnostopoulos:2010ux}
in the SSB of SO($D$).

\section{The Gaussian expansion method}
\label{sec:gaussian}

Since there are no quadratic terms in the actions
(\ref{eq:sb}) and (\ref{eq:sf}),
we cannot perform a perturbative expansion in the ordinary sense.
Finding the vacuum of this model is therefore a problem of 
solving a strongly coupled system.
It is known that a certain class of matrix models
can be solved exactly by using various large-$N$ techniques,
but the present model does not belong to 
such a category.\footnote{Note,
however, that the large-$N$ limit simplifies the calculation
in the Gaussian expansion method considerably
since it allows us to consider only the planar diagrams.
}
The use of the Gaussian expansion method 
in studying large-$N$ matrix quantum mechanics 
has been advocated by Kabat and Lifschytz \cite{Kabat:2000hp},
and various black hole physics of the dual geometry 
has been discussed \cite{blackholes}.
Applications to simplified versions of the type IIB matrix model
were pioneered by refs.\ \cite{Gauss_simpleIIB}.

The starting point of the Gaussian expansion method
is to introduce a Gaussian term $S_0$ and to rewrite the action 
$S=S_{\rm b}+S_{\rm f}$ as 
\begin{equation}
  S = (S_0 +  S) - S_0  \ .
\label{eq:shift_s}
\end{equation}
Then we can perform a perturbative expansion
regarding the first term $(S_0 +  S)$ as
the ``classical action'' and the second term
$(-S_0)$ as the ``one-loop counter term''.
The results at finite order depend, of course, on the choice
of the Gaussian term $S_0$, which contains many free parameters
in general.
However, it is known in various examples that there exists a region
of parameters, in which the results obtained at finite order
are almost constant.
Therefore, if we can identify this ``plateau region'',
we can make concrete predictions.
It should be emphasized that the method enables us to
obtain genuinely nonperturbative results,
although most of the tasks involved are
nothing more than perturbative calculations 
as emphasized in ref.\ \cite{Stevenson:1981vj}.

There are some cases in which one finds 
more than one plateau regions in the
parameter space.
In that case, each of them is considered to correspond to a local
minimum of the effective action, and
the plateau which gives the smallest free energy 
corresponds to the true vacuum. 
These statements have been confirmed explicitly
in exactly solvable matrix models \cite{Nishimura:2003gz}.

As the Gaussian action for the present model,
we consider the most general one that preserves 
the $\SU(N)$ symmetry.
Note, in particular, that 
we have to allow the Gaussian action 
to break the $\SO(10)$ symmetry
so that we can study the SSB of $\SO(10)$.
In practice we are going to restrict the parameter space
by imposing the $\SO(d)$ symmetry with $2 \le d \le 7$.
The plateau region identified for each $d$ corresponds
to a local minimum which breaks the $\SO(10)$ symmetry spontaneously.
By comparing the free energy,
we can determine which local minimum is actually the true vacuum.

Making use of the $\SO(10)$ symmetry of the model,
we can always bring the Gaussian action into the form
\begin{eqnarray}
  S_0 &=& S_{\rm 0b} + S_{\rm 0f} \ , 
\label{eq:gaussian} \\
  S_{\rm 0b} &=& \frac{N}{2} \sum_{\mu = 1}^{10} M_\mu \Tr (A_\mu)^2 \ , 
\label{eq:s0b} \\
  S_{\rm 0f} &=& \frac{N}{2} \sum_{\alpha,\beta=1}^{16} \mathcal{A}_{\alpha\beta} \Tr 
  ( \Psi_\alpha \Psi_\beta ) \ ,
\label{eq:s0f}
\end{eqnarray}
where $M_\mu$ and $\mathcal{A}_{\alpha\beta}$ are arbitrary parameters. 
The $16\times16$ complex matrix $\mathcal{A}_{\alpha\beta}$ 
can be expanded in term of the gamma 
matrices as
%
\begin{equation}
  \mathcal{A}_{\alpha\beta} 
  = \sum_{\mu,\nu,\rho = 1}^{10} \frac{i}{3!} m_{\mu\nu\rho} 
  (C \Gamma_{\mu} \Gamma_{\nu}^\dagger \Gamma_{\rho})_{\alpha\beta} \ ,
\label{eq:decomp_m}
\end{equation}
using a 3-form $m_{\mu\nu\rho}$.
We can then rewrite the partition function (\ref{eq:10dpf}) as
\begin{eqnarray}
  Z &=& Z_0 \,\langle e^{-(S - S_0)} \rangle_0 \ , 
\\
  Z_0 &=& \int \dd A \, \dd \Psi \, e^{-S_0} \ ,
\end{eqnarray}
where $\langle\,\cdot\,\rangle_0$ is a vacuum expectation value 
with respect to the partition function $Z_0$. 
{}From this we find
that the free energy $F= - \log Z$ can be expand as 
\begin{eqnarray}
  F &=& \sum_{k = 0}^\infty f_k \ , 
\nonumber \\
  f_0 &=& - \log Z_0 \ , 
\nonumber \\
  f_k &=& - \sum_{l = 0}^{k} 
  \frac{(-1)^{k-l}}{(k + l)!}\,{}_{k+l}{\rm C}_{k-l}
  \Bigl\langle (S_{\rm b} - S_0)^{k-l} (S_{\rm f})^{2l} 
  \Bigr\rangle_{\rm C,0} 
  \qquad \text{for $k\geq1$} \ ,
\label{eq:f_gem}
\end{eqnarray}
where the subscript `C' in $\langle\,\cdot\,\rangle_{\rm C,0}$ 
implies that the connected part is taken. 
The expansion is organized in such a way that
it corresponds to the loop expansion regarding the insertion
of the 2-point vertex $(-S_0)$ as a contribution from
the one-loop counterterm.
Similarly the expectation value of an observable $\mathcal{O}$ 
can be evaluated as
\begin{eqnarray}
  \langle \mathcal{O} \rangle &=& \langle \mathcal{O} \rangle_0
  + \sum_{k=1}^\infty O_k \ , 
\nonumber \\
  O_k &=& 
  \sum_{l = 0}^{k} \frac{(-1)^{k-l}}{(k + l)!}\,{}_{k+l}{\rm C}_{k-l}
  \langle 
    \mathcal{O}\,(S_{\rm b} - S_0)^{k-l}\,(S_{\rm f})^{2l} 
  \rangle_{\rm C,0} \  .
\label{eq:o_gem}
\end{eqnarray}

In practice we truncate the series expansion at some finite order.
Then the free energy (\ref{eq:f_gem}) and 
the observable (\ref{eq:o_gem}) 
depend on 
the free parameters $M_\mu$ and $\mathcal{A}_{\alpha\beta}$
in the Gaussian action.
%
We search for the values of parameters, at which
the free energy becomes stationary by solving 
the ``self-consistency equations''
\begin{equation}
  \frac{\partial}{\partial M_\mu} F = 0 \ , 
\qquad
  \frac{\partial}{\partial m_{\mu\nu\rho}} F = 0 \ ,
\label{eq:self-consistency}
\end{equation}
and estimate $F$ and $\langle\mathcal{O}\rangle$ 
at the solutions.
As we increase the order of the expansion, the number
of solutions increases. If we find that there are many solutions
close to each other in the parameter space which give similar
results for the free energy and the observables, we may identify
the region as a plateau.


In actual calculation 
it is convenient to derive the series expansion
(\ref{eq:f_gem}) in the following way.
First we consider the action 
\begin{equation}
  \tilde{S} = S_0 + \epsilon S_{\rm b} + \sqrt{\epsilon} S_{\rm f} \ ,
\label{eq:imfa_s}
\end{equation}
and the partition function 
\begin{equation}
  \tilde{Z} = \int \dd A \, \dd \Psi \, e^{-\tilde{S}} \ , 
\label{eq:imfa_pf}
\end{equation}
where $\epsilon$ is a fictitious expansion parameter.
Next we calculate the free energy 
in the $\epsilon$-expansion as
\begin{equation}
  \tilde{F} = - \log \tilde{Z} = \sum_{k=0}^{\infty} \epsilon^k 
\tilde{f}_{k} \ .
\label{eq:imfa_f}
\end{equation}
Each term $\tilde{f}_{k}$ depends on
the parameters $M_\mu$ and $m_{\mu\nu\lambda}$ 
in the Gaussian action $S_0$.
Then we substitute these parameters as
\begin{equation}
  M_\mu \to (1-\epsilon)\,M_\mu \ ,
\qquad
  m_{\mu\nu\rho} \to (1-\epsilon)\,m_{\mu\nu\rho} \ ,
\label{eq:imfa_improve}
\end{equation}
reorganize the series with respect to $\epsilon$,
and set $\epsilon$ to~1. 
In this way we reproduce the expression (\ref{eq:f_gem}). 
The action (\ref{eq:imfa_s})
is introduced to obtain the ordinary perturbation theory for 
the first term in (\ref{eq:shift_s}),
and the final step (\ref{eq:imfa_improve})
corresponds to taking account of the second term
in (\ref{eq:shift_s}) as the one-loop counter term.
The main task is to obtain the series (\ref{eq:imfa_f}),
which is nothing more than what is required for 
ordinary perturbation theory.\footnote{Further simplification is 
possible by exploiting the fact that the free energy $F$ is related 
to the two-particle irreducible (2PI) free energy
through the Legendre transformation \cite{Kawai:2002jk}. 
The number of Feynman diagrams decreases considerably
by the restriction to 2PI diagrams.
While we do not use this technique for the 3rd order calculations
in the present work,
it would be crucial in performing higher order calculations.
}
We use a similar procedure to obtain the expansion (\ref{eq:o_gem})
for the observables.

Since we are interested in the large-$N$ limit, 
we list up Feynman diagrams in the double-line notation,
and keep only the planar diagrams when evaluating
the free energy (\ref{eq:f_gem}) and 
the observable (\ref{eq:o_gem}).





\TABLE[t]{%
{\small
\begin{tabular}{|c|l|c|}
\hline
ansatz & \hspace{5mm}discrete symmetry for shrunken directions  & symbol \\ \hline
SO(7) & & $\circ$ \\ \hline
SO(6) & $ \mathbb{Z}_{4(7,8,9,10|1)} $ & $\circ$ \\ \cline{2-3}
      & $ \mathbb{Z}_{3(8,9,10)} $ & $\bullet$ \\ \cline{2-3}
      & $ \mathbb{Z}_{2(7,8|1)} $ & $\tri$ \\ \hline
SO(5) & $ \mathbb{Z}_{5(6,7,8,9,10)} $ & $\circ$ \\ \cline{2-3}
      & $ \mathbb{Z}_{4(6,7,8,9|1)} $ & $\bullet$ \\ \cline{2-3}
      & $ \mathbb{Z}_{4(6,7,8,9|10)} $ & $\tri$ \\ \cline{2-3}
      & $ \mathbb{Z}_{2(6,7|1)}\times\mathbb{Z}_{3(8,9,10)} $ & $\btri$ \\ \cline{2-3}
      & $ \mathbb{Z}_{2(6,7|8,9,10)}\times\mathbb{Z}_{3(8,9,10)} $ & $\dtri$ \\ \cline{2-3}
      & $ \mathbb{Z}_{2(6,7|1)}\times\mathbb{Z}_{2(8,9|1)} $ & $\dbtri$ \\ \cline{2-3}
      & $ \mathbb{Z}_{2(6,7|10)}\times\mathbb{Z}_{2(8,9|10)}
\times\mathbb{Z}_{2(\{6,7\},\{8,9\}|1,10)} $ & $\sq$ \\ \hline
SO(4) & $ \mathbb{Z}_{2(5,6|1)}\times\mathbb{Z}_{2(7,8|1)}
\times\mathbb{Z}_{2(9,10|1)} $ & $\circ$ \\ \cline{2-3}
      & $ \mathbb{Z}_{3(5,6,7)}\times\mathbb{Z}_{3(8,9,10)}
\times\mathbb{Z}_{2(\{5,6,7\},\{8,9,10\}|1)}$ & $\bullet$ \\ \cline{2-3}
      & $ \mathbb{Z}_{4(5,6,7,8|1)}\times\mathbb{Z}_{2(9,10|1)} $ & $\tri$ \\ \cline{2-3}
      & $ \mathbb{Z}_{5(6,7,8,9,10)}\times\mathbb{Z}_{2(-|1,6,7,8,9,10)} $ & $\btri$ \\ \cline{2-3}
      & $ \mathbb{Z}_{3(5,6,7)}\times\mathbb{Z}_{3(8,9,10)}\times\mathbb{Z}_{2(-|1,8,9,10)} $ & $\dtri$ \\ \hline
SO(3) & $ \mathbb{Z}_{2(4,5|1)}\times\mathbb{Z}_{2(6,7|1)}
\times\mathbb{Z}_{2(8,9|1)}\times\mathbb{Z}_{3(\{4,5\},\{6,7\},\{8,9\})}$ & $\circ$ \\ \cline{2-3}
      & $ \mathbb{Z}_{2(4,5|10)}\times\mathbb{Z}_{2(6,7|10)}
\times\mathbb{Z}_{2(8,9|10)}$ & $\bullet$ \\
      & $ \quad \times\mathbb{Z}_{3(\{4,5\},\{6,7\},\{8,9\})}\times\mathbb{Z}_{2(\{4,5\},\{6,7\})}$ & \\ \cline{2-3}
      & $ \mathbb{Z}_{2(4,5|10)}\times\mathbb{Z}_{2(6,7|10)}
\times\mathbb{Z}_{2(8,9|10)} $ & $\tri$ \\
      & $ \quad \times\mathbb{Z}_{3(\{4,5\},\{6,7\},\{8,9\})}\times\mathbb{Z}_{2(-|1,10)} $ & \\ \cline{2-3}
      & $ \mathbb{Z}_{3(4,5,6)}\times\mathbb{Z}_{3(7,8,9)}
\times\mathbb{Z}_{2(\{4,5,6\},\{7,8,9\}|1)}\times\mathbb{Z}_{2(-|1,10)} $ & $\btri$ \\ \cline{2-3}
      & $ \mathbb{Z}_{3(4,5,6)}\times\mathbb{Z}_{2(7,8|1)}\times\mathbb{Z}_{2(9,10|1)}\times\mathbb{Z}_{2(\{7,8\},\{9,10\})} $ & $\dtri$ \\ \cline{2-3}
      & $ \mathbb{Z}_{3(4,5,6)}\times\mathbb{Z}_{4(7,8,9,10|1)}\times\mathbb{Z}_{2(-|7,8,9,10)} $ & $\dbtri$ \\ \cline{2-3}
      & $ \mathbb{Z}_{2(4,5|1)}\times\mathbb{Z}_{5(6,7,8,9,10)}\times\mathbb{Z}_{2(-|1,6,7,8,9,10)} $ & $\sq$ \\ \cline{2-3}
      & $ \mathbb{Z}_{6(5,6,7,8,9,10|1)}\times\mathbb{Z}_{2(-|5,6,7,8,9,10)} $ & $\bsq$ \\ \cline{2-3}
      & $ \mathbb{Z}_{2(4,5|1)}\times\mathbb{Z}_{5(6,7,8,9,10|4,5)} $ & \\ \hline
SO(2) & $ \mathbb{Z}_{2(3,4|1)}\times\mathbb{Z}_{2(5,6|1)}
\times\mathbb{Z}_{2(7,8|1)}\times\mathbb{Z}_{2(9,10|1)} $ & $\circ$ \\
      & $ \quad
\times\mathbb{Z}_{4(\{3,4\},\{5,6\},\{7,8\},\{9,10\})} $ & $$ \\ \cline{2-3}
      & $ \mathbb{Z}_{4(3,4,5,6|1)}\times\mathbb{Z}_{4(7,8,9,10|1)}
\times\mathbb{Z}_{2(\{3,4,5,6\},\{7,8,9,10\})} $ & $\bullet$ \\ \cline{2-3}
      & $ \mathbb{Z}_{3(3,4,5)}\times\mathbb{Z}_{3(6,7,8)}
\times\mathbb{Z}_{2(\{3,4,5\},\{6,7,8\}|1)} $ & $\tri$ \\
      & $ \quad \times\mathbb{Z}_{2(9,10|1)}\times\mathbb{Z}_{2(-|3,4,5,6,7,8)} $ & \\ \cline{2-3}
      & $ \mathbb{Z}_{3(3,4,5)}\times\mathbb{Z}_{3(6,7,8)}
\times\mathbb{Z}_{2(\{3,4,5\},\{6,7,8\}|1)} $ & $\btri$ \\
      & $ \quad \times\mathbb{Z}_{2(9,10|1)}\times\mathbb{Z}_{2(-|9,10)} $ & \\ \hline
\end{tabular}

}
\caption{%
The list of the $\SO(d)$ ansatz ($d=2,\cdots,7$), which leave us
with not more than 5 parameters in the Gaussian action.
The symbols shown in the right-most column
are used 
in figs.~\ref{fig:f}, \ref{fig:f2}, \ref{fig:R} and \ref{fig:r}
to represent the solutions to the self-consistency equations.
For the 
$\SO(3)\times\mathbb{Z}_{2(4,5|1)}\times\mathbb{Z}_{5(6,7,8,9,10|4,5)}$
ansatz,
we find no solutions
in the range $-2<f<7$ displayed in fig.~\ref{fig:f}, 
and hence we do not assign any symbol.
}
\label{tbl:ansatz}
}

There are many free parameters in the Gaussian action; \emph{i.e.,}
we get
10 from $M_\mu$ and 120 from $m_{\mu\nu\rho}$.
There are as many self-consistency equations as these parameters.
Unfortunately it seems impossible to solve them
in full generality.
However, it is reasonable to expect that some subgroup of SO(10)
symmetry such as SO($d$) with $2\leq d \leq 7$ remains unbroken.
This allows us to impose the corresponding symmetry on the 
Gaussian action, and hence to reduce the number of parameters.

In order to study 
the SO($d$) symmetric vacuum, we impose the SO($d$) symmetry 
on the Gaussian action.
For $d \ge 4$, this leads to $M_1 = \cdots = M_d$ and
$m_{\mu\nu\rho}=0$ 
unless $\mu$, $\nu$ and $\rho$ are
three different numbers in $\{ d+1, \cdots , 10 \}$.
Therefore 
we obtain
$(11-d)$ parameters from $M_\mu$ and
${}_{10-d} \,  {\rm C \, }_3$ parameters from $m_{\mu\nu\rho}$.
For $d=3$, $m_{123}$ can also be non-zero, and
for $d=2$, $m_{12 \mu}$ with $\mu=3 , 4 , \cdots , 10 $ 
can also be non-zero.
Thus the total number of parameters is
5, 9, 16, 27, 44, 73 for $d=7,6,5,4,3,2$, respectively.
For
$d \ge 8$, the SO($d$) symmetry enforces $m_{\mu\nu\rho}=0$,
which makes the Gaussian expansion ill-defined. However, 
the ${\rm SO}(d)$ symmetric vacua with $d \ge 8$ are expected to
exist, and they should be realized as a plateau with
the SO($d$) ansatz ($d\le 7$).
See ref.~\cite{Aoyama:2006je} for an explicit example of this claim
for the ${\rm SO}(10)$ symmetric vacuum. 

It turned out that solving the self-consistency equations 
(\ref{eq:self-consistency})
is possible
for the number of parameters not more than 5.
Except for $d=7$, we therefore 
impose extra discrete symmetries ($\Sigma_d$)
in the shrunken directions $x_{d+1},\cdots,x_{10}$,
where we require that ${\rm SO}(d) \times \Sigma_d$ 
is a subgroup of SO(10).
Let us introduce the following notations.
First,
$\mathbb{Z}_{p(i_1,\cdots,i_p)}$ represents the group of
cyclic permutations acting on $x_{i_1} , \cdots , x_{i_p}$.
Similarly, 
$\mathbb{Z}_{p(\{i_1,j_1\},\cdots,\{i_p,j_p\})}$ represents 
the group of cyclic permutations acting on the sets
$\{x_{i_1}, x_{j_1}\},\cdots,\{x_{i_p},x_{j_p}\}$.
We also combine them with a reflection.
For instance, the symbol
$\mathbb{Z}_{p(i_1,\cdots,i_p|j_1,\cdots,j_q)}$
implies that we make a reflection 
$(x_{j_1},\cdots,x_{j_q}) \mapsto (- x_{j_1},\cdots,- x_{j_q})$
together with a cyclic permutation of $x_{i_1}\cdots,x_{i_p}$.
The symbol $\mathbb{Z}_{2(-|j_1,\cdots,j_q)}$ 
implies that we make a reflection without any cyclic permutations.
Note that the permutation is odd
if the integer $p$ is even and 
the number of elements in the set to be permuted is odd.
In that case, 
we have
to combine it with a reflection 
in an odd number of directions in order to make the imposed
symmetry an element of ${\rm SO}(10)$.
Table \ref{tbl:ansatz} shows the list of the ansatz that
leave us with not more than 5 parameters.
In the previous works \cite{Nishimura:2001sx,%
Kawai:2002jk,Kawai:2002ub,Aoyama:2006di,Aoyama:2006rk,Aoyama:2006je}
one particular extra symmetry was chosen for each $d$
so that the number of parameters is reduced to 3.
All the extra symmetries used in the literature
are included in the list of ansatz we study.
See appendix \ref{sec:app_ansatz} for more detail.

\section{Results}
\label{sec:result}

\begin{figure}[H]
\begin{center}
\includegraphics[width=0.80 \linewidth]{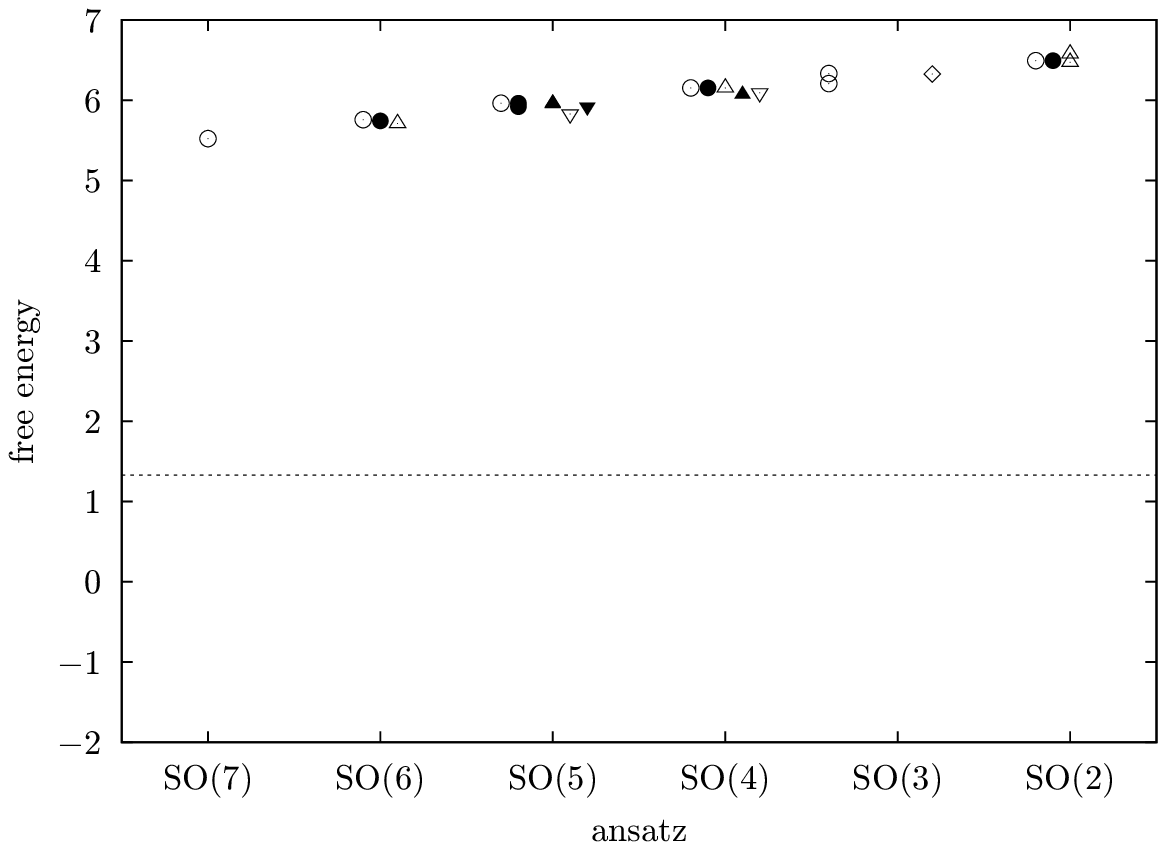}
\includegraphics[width=0.80 \linewidth]{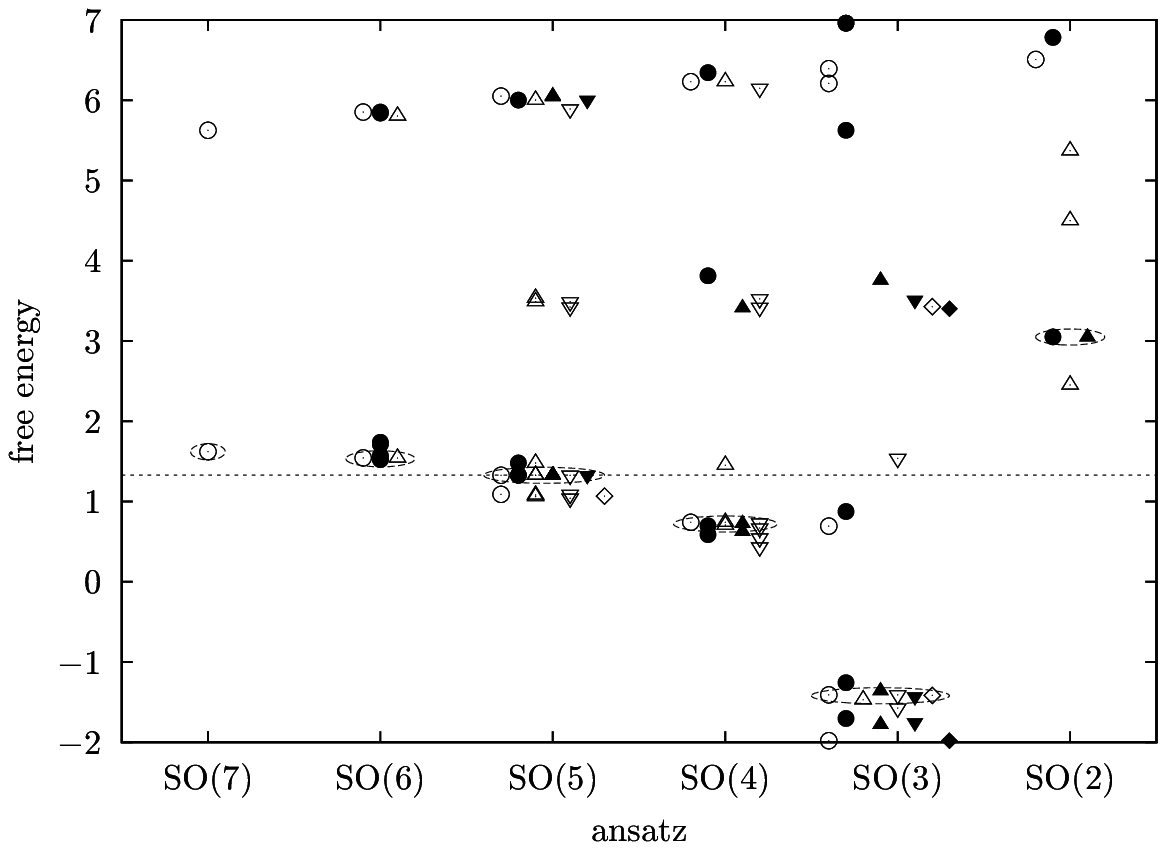}
\end{center}
\caption{%
The free energy density (\ref{eq:fedensity})
evaluated at the solutions to the self-consistency equations 
at the orders 1 (Top) and 3 (Bottom).
Each symbol represents the largest symmetry in Table \ref{tbl:ansatz}
that the solution has.
The horizontal line represents the value $\log 8 - \frac{3}{4} = 1.32944...$  
obtained from the KNS conjecture. The data points surrounded by
the dashed lines correspond to the ``physical solutions'',
which indicate the plateau region.}
\label{fig:f}
\end{figure}

For each ansatz,
we first obtain the free energy 
up to the third order
as a function of the free parameters in the Gaussian action.
By differentiating the free energy with respect to the
free parameters, we obtain
the self-consistency equations,
which we solve numerically by Mathematica.
The free energy evaluated at each solution is
plotted in fig.\ \ref{fig:f}
for the SO($d$) ansatz ($2\leq d\leq 7$) described 
in the previous section.
More precisely, we actually plot ``the free energy density'' defined as
\begin{equation}
  f = 
  \lim_{N\to\infty} \left\{
  \frac{F}{N^2 - 1} 
  - \left( - 3 \log N \right) 
  \right\} \ ,
\quad
\mbox{where~}F=-\log Z \ .
\label{eq:fedensity}
\end{equation}


\FIGURE[H]{%
\epsfig{file=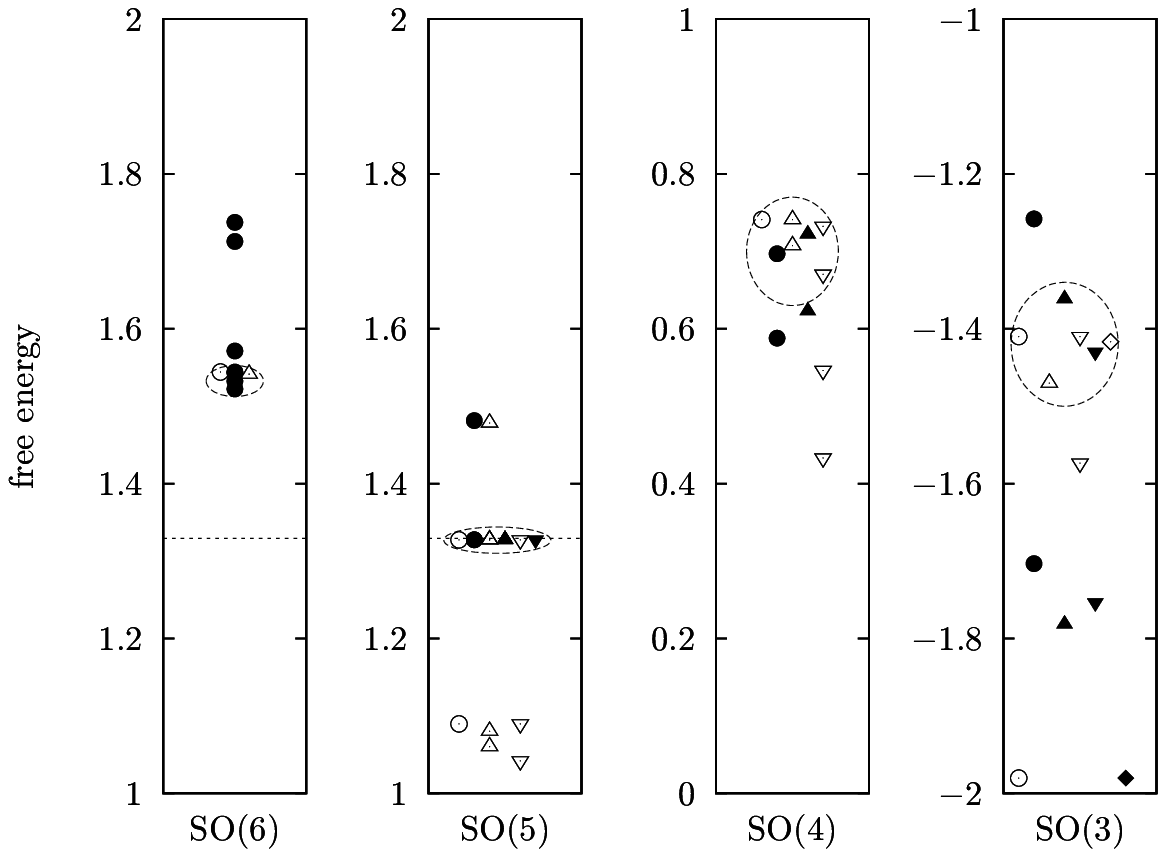,width=.8\textwidth}
\caption{%
The zoom-up of fig.\ \ref{fig:f} (Bottom) for $d=6,5,4,3$
near the ``physical solutions'', which are 
surrounded by the dashed line.
}
\label{fig:f2}
}

\FIGURE[H]{%
\epsfig{file=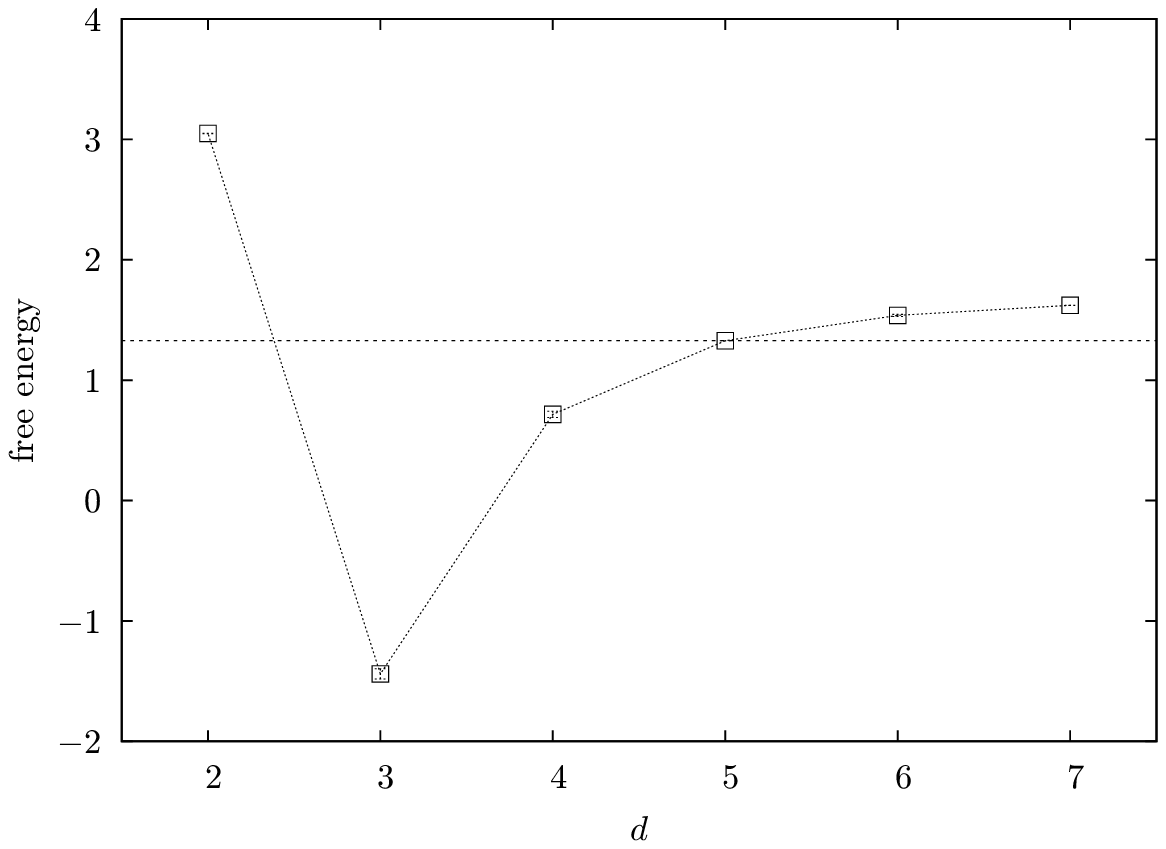,width=.8\textwidth}
\caption{%
The free energy density
averaged over the ``physical solutions'' for each $d$
is plotted against $d$. 
We put error bars representing the mean square error
when there are more than one physical solutions.
The horizontal
line represents the KNS value $f=\log 8 - \frac{3}{4} = 1.32944\ldots$, 
and the dotted line connecting
the data points is drawn to guide the eye.}
\label{fig:free-summary}
}

\newpage


The horizontal line $\log 8 -\frac{3}{4}=1.32944...$
represents the result
obtained in appendix \ref{sec:KNSresult} 
from the 
analytic formula of the partition function conjectured 
by Krauth, Nicolai and Staudacher (KNS) \cite{Krauth:1998xh}.
%
At order 3, the ``physical solutions'' identified 
for each $d$ as described below
are surrounded by a dashed line.
In fig.\ \ref{fig:f2} we zoom up the region 
containing these solutions
for $ d = 6,5,4,3$,
which gives the free energy density
$f \sim  1.5,\ 1.3,\ 0.7,\ -1.4$, respectively.

In fig.~\ref{fig:free-summary} we show
the free energy density obtained by averaging over the physical solutions
for each $d$ at order 3.
We put error bars representing the mean square error
when there are more than one physical solutions.
The result decreases monotonically as $d$ decreases from 7 to 3,
and it becomes much larger for $d=2$.
Thus, the $\SO(3)$ symmetric vacuum gives the smallest free energy density.
The $d$-dependence of the free energy density
is quite analogous to the one observed in the six-dimensional
case \cite{Aoyama:2010ry}.
There the value of the free energy tends to decrease
slightly as one goes from order 3 to order 5.
Considering such artifacts due to truncation,
we speculate that the KNS conjecture actually 
refers to the partition function for the ${\rm SO}(10)$ symmetric vacuum.

Let us then explain how we identify the ``physical solutions''.
For that we also refer to the results for the extent of space-time.
In appendix \ref{sec:app_ansatz} we give the explicit form of 
$\langle T_{\mu\nu} \rangle$ for each ansatz. 
Let us note that it is not diagonal in general,
and therefore we need to diagonalize it in order to obtain 
the eigenvalues $\langle \lambda_j  \rangle$.
Then we find that some solutions give smaller values in the
directions involved in the preserved $\SO(d)$ symmetry
than in the remaining directions.
Such solutions are not shown in the figures, and will not be considered
in what follows. 
For the $\SO(d)$ ansatz, the $d$ large eigenvalues 
$\langle \lambda_j  \rangle$ $(1 \le j \le d)$
are equal due to the imposed $\SO(d)$ symmetry,
and we denote the corresponding value as $R^2$. 
The remaining $(10-d)$ eigenvalues for each
solution turn out to be quite close to each other
and we denote the mean value as $r^2$. 
In figs.~\ref{fig:R} and \ref{fig:r},
we plot the values of $R^2$ and $r^2$ evaluated at the solutions
for each ansatz.

At order 3, we find a set of solutions for $3 \le d  \le 6$
giving similar values for the free energy density $f$ 
and the extent of space-time $R^2$ and $r^2$.
We consider this as the concentration of solutions \cite{Kawai:2002jk},
which indicates a plateau region in the space of parameters
explained in section \ref{sec:gaussian}.
Thus we can pick up the ``physical solutions'' for $3 \le d  \le 6$
without much ambiguity.
For the $\SO(2)$ ansatz, we find two solutions close to each other
with $f \sim 3.0$, $R^2 \sim 3.6$ and $r^2 \sim 0.11$, which we identify 
as the physical solutions.
For the $\SO(7)$ ansatz, we obtain only two solutions at order 3,
which are not close to each other.
From fig.~\ref{fig:f} (Bottom), however, 
it looks reasonable to
identify the one with smaller free energy density
as the physical solution.


In fig.~\ref{fig:rr} we plot the result for $R^2$ and $r^2$
averaged over all the physical solutions for each $d$. 
We put error bars representing the mean square error
when there are more than one physical solutions.
We find that $r^2$ stays almost constant at $r^2 = 0.1 \sim 0.15$,
which seems to be universal for all the SO($d$) symmetric
vacua with $2 \le d \le 7$.
On the other hand, the results for $R^2$
are found to
be larger for smaller $d$.
Let us test whether this behavior is consistent with the 
constant-volume property \cite{Aoyama:2010ry}, which 
is given by $R^d \, \tilde{r}^{10-d} \approx \ell^{10}$,
where 
$\tilde{r}^2$ represents the universal ``compactification'' scale,
which we leave as a free parameter in the present analysis.

\newpage

\FIGURE[H]{%
\begin{tabular}{c}
\epsfig{file=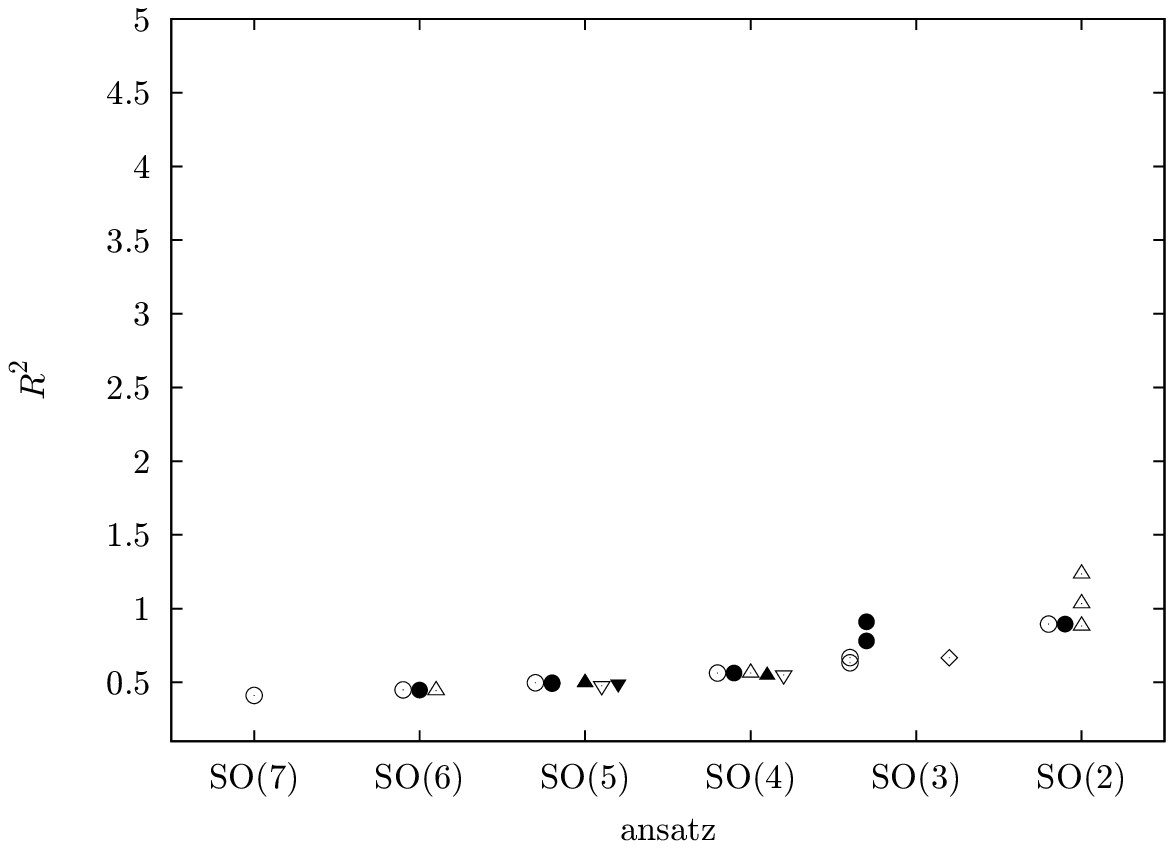,width=.9\textwidth} \\
\epsfig{file=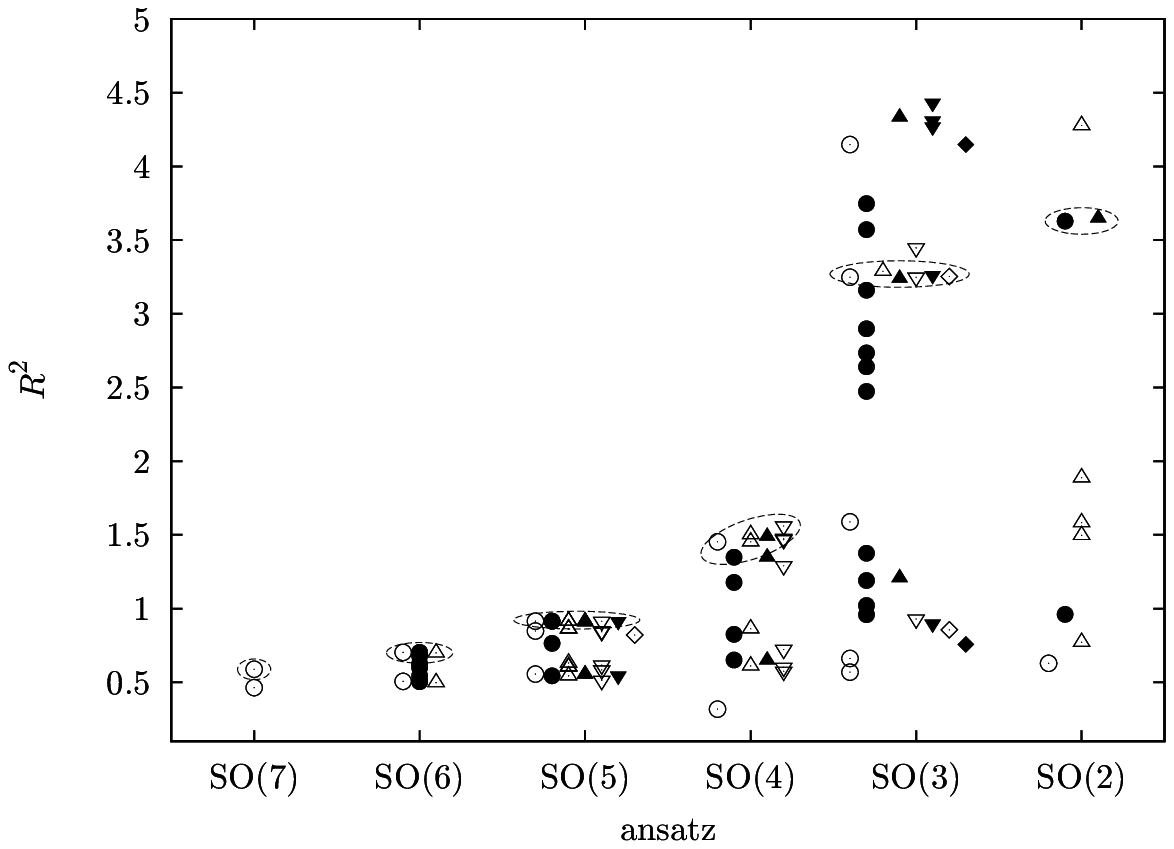,width=.9\textwidth}
\end{tabular}
\caption{%
The extent of space-time $R^2$ in the extended directions
evaluated at the solutions to the self-consistency equations
at the first order (Top) and the third order (Bottom).
Each symbol represents the largest symmetry in Table \ref{tbl:ansatz}
that the solution has.
The data points surrounded by
the dashed lines correspond to the ``physical solutions'',
which indicate the plateau region.
}
\label{fig:R}
}

\newpage

\FIGURE[H]{%
\begin{tabular}{c}
\epsfig{file=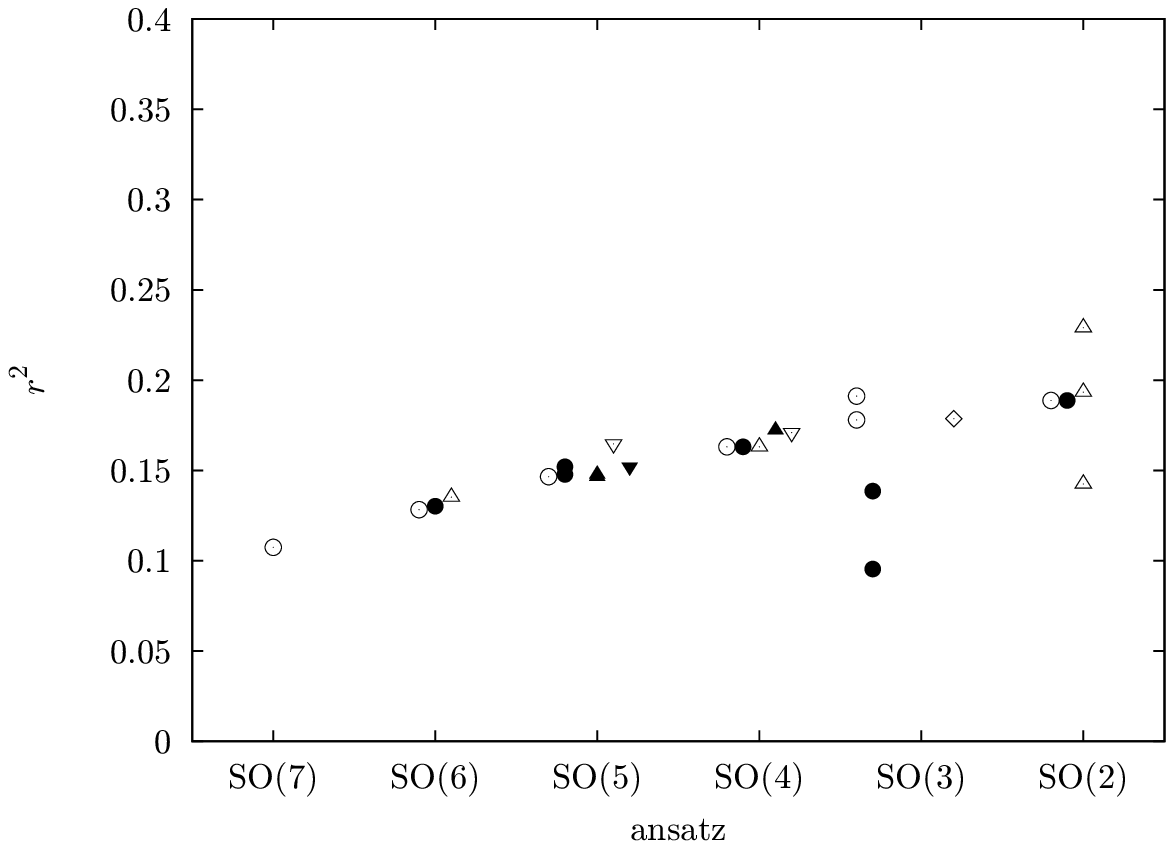,width=0.9\textwidth} \\
\epsfig{file=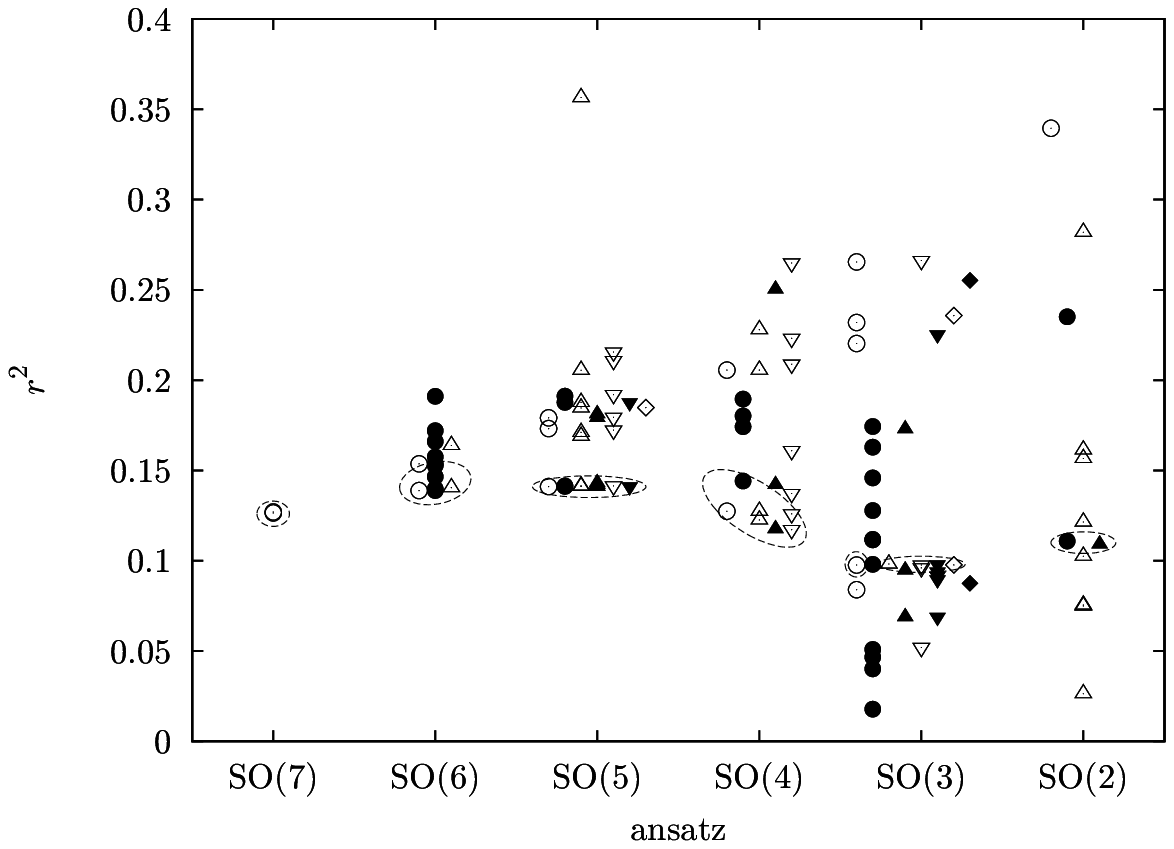,width=0.9\textwidth}
\end{tabular}
\caption{%
The extent of space-time $r^2$ in the shrunken directions
evaluated at the solutions to the self-consistency equations
at the first order (Top) and the third order (Bottom).
Each symbol represents the largest symmetry in Table \ref{tbl:ansatz}
that the solution has.
The data points surrounded by
the dashed lines correspond to the ``physical solutions'',
which indicate the plateau region.
}
\label{fig:r}
}

\newpage

\FIGURE[H]{%
\epsfig{file=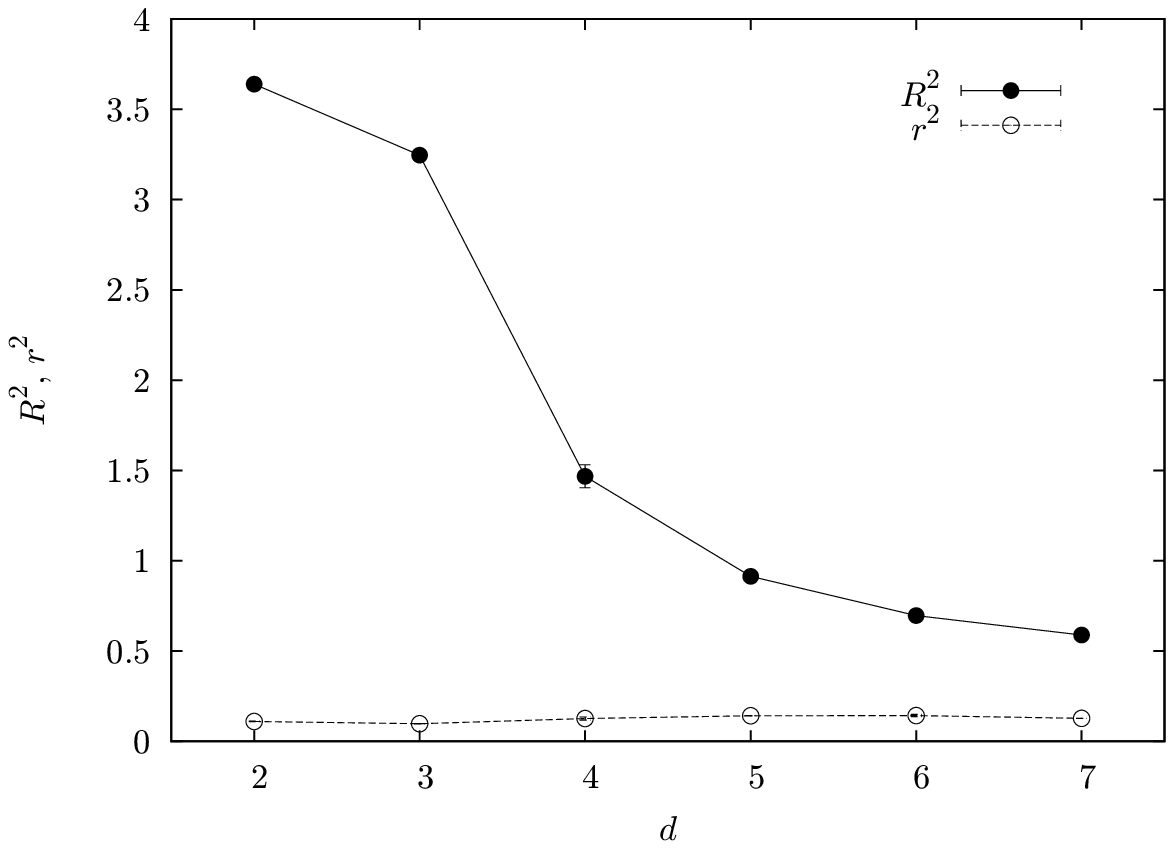,width=.8\textwidth}
\caption{%
The extent of space-time $R^2$ and $r^2$ 
in the extended
and shrunken directions, respectively, 
are plotted against $d$ after taking the average
over the ``physical solutions'' for each $d$.
We put error bars representing the mean square error
when there are more than one physical solutions.
The solid and dashed lines connecting
the data points are drawn to guide the eye.
}
\label{fig:rr}
}

\FIGURE[H]{%
\epsfig{file=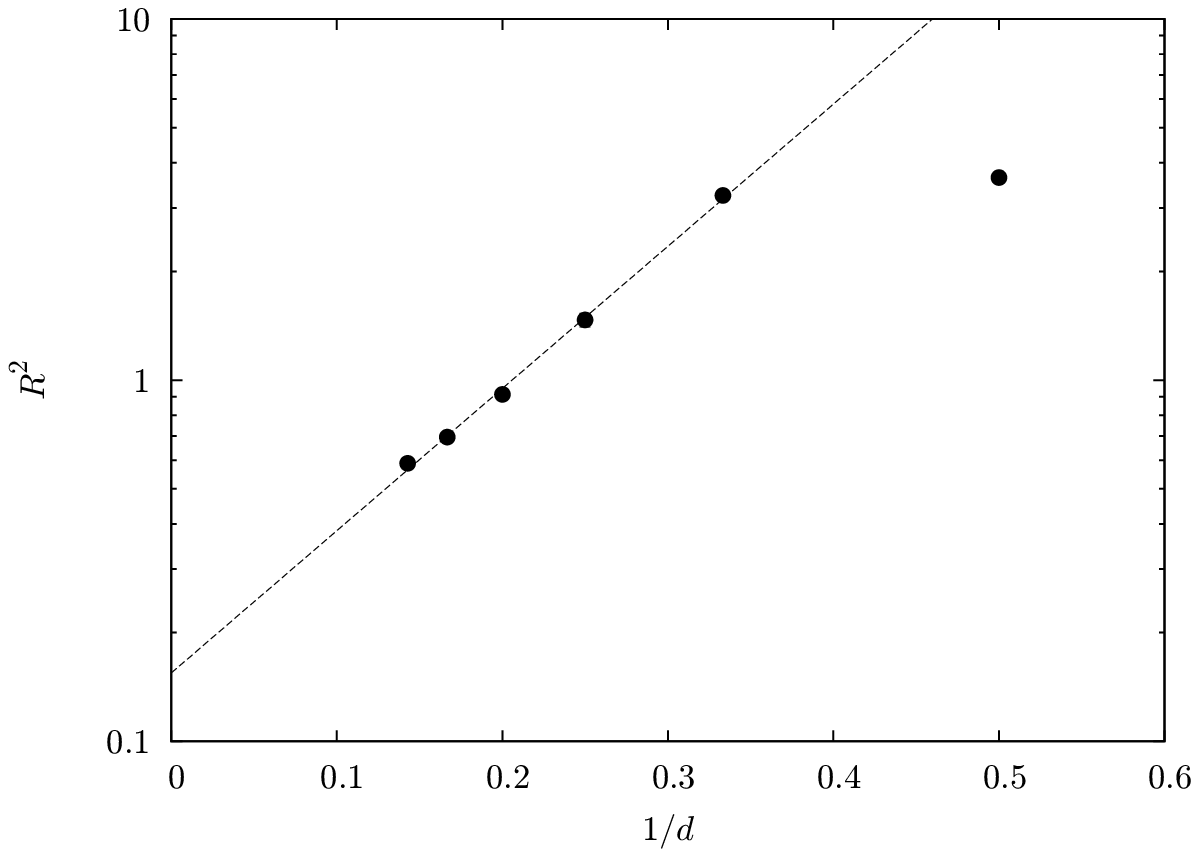,width=.8\textwidth}
\caption{
The extent of space-time $R^2$ in the extended directions
shown in fig.~\ref{fig:rr} is plotted in the log scale
against $1/d$.
The straight dashed line represents
a fit to the behavior (\ref{logR-formula}) 
excluding $d=2$,
from which we obtain $\tilde{r}^2=0.155$ and $\ell^2 = 0.383$.
}
\label{fig:ext-R}
}

\newpage

{}From the constant-volume property, 
we obtain
\beq
  \log R^2 \approx 
 \frac{10}{d} \log \left( \frac{\ell^2}{\tilde{r}^2} \right)
+ \log \tilde{r}^2 \ .
\label{logR-formula}
\eeq
In fig.~\ref{fig:ext-R} we therefore 
plot $R^2$ in the log scale against $1/d$.
Indeed we find that the results
can be fitted
to a straight line except for $d = 2$.
From the fit, 
we obtain $\ell^2 = 0.383$ and $\tilde{r}^2 = 0.155$.
Note that the obtained value for $\tilde{r}^2$
is consistent with $r^2$
obtained directly by the Gaussian expansion method
as the extent of space-time in the shrunken directions.
The obtained value for $\ell^2$ is reproduced by Monte Carlo
simulation \cite{AAN}.


\section{Summary and discussions}
\label{sec:discussion}

In this paper 
we have applied the Gaussian expansion method
to the type IIB matrix model,
which was conjectured to be a nonperturbative formulation
of type IIB superstring theory in ten dimensions.
In particular, we have 
investigated the dynamical properties of the model
associated with the SSB of SO(10),
which was speculated to realize the dynamical compactification.
Unlike previous works, we studied
the SO($d$) symmetric vacua for $2 \le d \le 7$ systematically.
Moreover, we were able to observe clear indication
of plateaus, which is crucial for the method to be reliable.
This was made possible by allowing various extra symmetries
in the shrunken directions as suggested by the success
of a similar work on the six-dimensional version of the model.

Indeed our results bear surprising similarity to
the results for the 6d case.
The free energy was found to decrease as $d$ is lowered
until one reaches $d=3$, and then it becomes much larger for $d=2$.
This implies that the SO(10) symmetry is spontaneously broken down 
to SO(3). The ``compactification'' scale 
is universal for all $d$.
The extent of space-time obeys 
the constant-volume property for $d \ge 3 $,
while the volume for $d=2$ is a bit smaller.
All these properties are common to the 6d and 10d cases,
and they can be understood intuitively as follows.
(See ref.~\cite{Aoyama:2010ry} for more detailed discussions.)

First of all, the long-distance effective theory of these
models can be obtained by integrating out the off-diagonal
elements of the matrices perturbatively \cite{Aoki:1998vn,Ambjorn:2000dx}.
One then finds that the effective theory for the diagonal elements
is described by a branched-polymer-like system,
where the diagonal elements of the bosonic matrices $A_\mu$
are identified with the coordinates of the vertices in the polymer.
This naturally explains the constant-volume property since
the branched polymer tends to occupy a fixed volume
for entropic reasons.
The anomaly for $d=2$ can be understood as a consequence of the fact
that the fermion determinant disfavors two-dimensional
configurations \cite{NV}.
The volume that appears in the description of
the constant-volume property can be calculated by Monte Carlo 
simulation,
and it agrees quite well with the value suggested by the Gaussian
expansion method for both 6d and 10d cases \cite{AAN}.

The driving force for the SSB comes from the 
phase of the fermion determinant \cite{NV}
as has been clearly demonstrated in 
a simplified non-supersymmetric model \cite{Nishimura:2001sq},
which was studied by both the Gaussian expansion method 
\cite{Nishimura:2004ts} and 
a Monte Carlo method \cite{Anagnostopoulos:2010ux},
giving consistent results.
There the free energy was determined by subtle competition
between the effect of the phase, which favors smaller $d$,
and the dynamics of the phase-quenched model, which favors larger $d$.
In the present supersymmetric matrix models,
the latter effect is suppressed by $1/N$ \cite{AAN,sign},
and the free energy
is determined essentially by the effect of the phase.
This explains the monotonic decrease of the free energy for smaller $d$.
The anomaly for $d=2$ can be understood again from the property of 
the fermion determinant.

The extent in the shrunken directions is determined by the
fluctuation of the off-diagonal elements since the 
diagonal elements cannot fluctuate in the shrunken directions
due to the strong effect of the phase.
At the leading order of the long-distance approximation,
the dynamics of the off-diagonal elements is given by the Gaussian
action, which couples weakly to the dynamics of the diagonal elements.
This explains the universal compactification scale.

Since our results are obtained at the 3rd order
of the Gaussian expansion,
it remains to be seen whether the main conclusions change qualitatively
at higher orders.
In fact the 6d case has been studied both at the 3rd and 5th orders.
The results
did change slightly, but the qualitative features seemed to be robust.
It is nonetheless desirable to perform the 5th order calculations
in the 10d case as well.
It would be also important to perform Monte Carlo 
studies analogous to \cite{Anagnostopoulos:2010ux} and 
to see whether the properties suggested by our work can be 
reproduced.

Our calculations clearly demonstrate
the SSB of SO(10),
which is itself an interesting dynamical property of 
the type IIB matrix model.
However, in view of the scenario for dynamical compactification,
there are two problems.
One is that it seems more likely that
the minimum of free energy occurs at $d=3$ instead of $d=4$,
at least within the approximation of the present work.
The other is that the ratio of the extent of space-time 
in the extended directions and that in the shrunken directions 
is finite for all $d$,
as suggested by the constant volume property and the 
universal compactification scale.
We consider that 
the nontrivial dynamics of the type IIB matrix model suggested
by the present work confirms the validity of the basic idea
to use matrices as the microscopic degrees of freedom 
in formulating superstring theory nonperturbatively.
The detailed properties of the space-time emerging from the model,
however, suggest that it still lacks some important ingredient
in order to be capable of describing our real world.

\acknowledgments
We would like to thank Hajime Aoki, Tatsumi Aoyama,
Takehiro Azuma, Masanori Hanada, Satoshi Iso, Hikaru Kawai,
Yoshihisa Kitazawa and Asato Tsuchiya for discussions.
The present work is supported in part by Grant-in-Aid 
for Scientific Research 
(No.\ 19340066 and 20540286 for J.N. and No.\ 21540290 for F.S.)
from Japan Society for the Promotion of Science.


\appendix

\section{Details of the ansatz}
\label{sec:app_ansatz}

In this section we describe the details of the ansatz,
which are listed in Table \ref{tbl:ansatz}.
We exhaust all the symmetries ${\rm SO}(d) \times \Sigma_d$
with $2 \le d  \le 7$, which reduce the number of 
parameters in the Gaussian action to 5 or less.
We also describe the general form 
of the moment of inertia tensor (\ref{eq:tmunu})
for each ansatz. 

\paragraph{$\SO(7)$ ansatz}

\begin{itemize}

\item SO(7) 

5 parameters ($M,M_8,M_9,M_{10},\tilde{m}$)
\beqa
  M_\mu = (\underbrace{M,\cdots,M}_7,M_8,M_9,M_{10}) \ , \quad
  m_{8,9,10} = \tilde{m} \ , \quad 
  \text{and zero otherwise.}
\eeqa
The moment of inertia tensor takes the form
\beq
  \langle T_{\mu\nu} \rangle = \left( 
  \begin{array}{@{\,}c|ccc@{\,}}
  {\rm SO(7) \, part} & & & \\ \hline
  & C & \times & \times \\
  & \times & C' & \times \\
  & \times & \times & C''
  \end{array}
\right) .
\eeq
Here and henceforth ``$\times$'' represents a component which is
found to vanish
up to order 3 by explicit calculation,
although there is no such symmetry that
enforces it.

By further imposing the symmetry $\mathbb{Z}_{3(8,9,10)}$,
which enforces 
$M_8 = M_9 = M_{10}$,
one obtains the ``$\SO(7)$ ansatz'' used in
refs.~\cite{Nishimura:2001sx,%
Kawai:2002jk,Kawai:2002ub,Aoyama:2006di,Aoyama:2006rk,Aoyama:2006je}.
\end{itemize}

\paragraph{$\SO(6)$ ansatz}

\begin{itemize}

\item $\SO(6)\times\mathbb{Z}_{4(7,8,9,10|1)}$

3 parameters ($M,\tilde{M},\tilde{m}$)
\begin{gather}
M_\mu = (\underbrace{M,\cdots,M}_6,\tilde{M},\tilde{M},\tilde{M},\tilde{M}) 
  \ , \nonumber \\
 m_{7,8,9} = m_{8,9,10} = m_{7,9,10} = m_{7,8,10} = \tilde{m} \ , \quad
  \text{and zero otherwise.}
\end{gather}
The moment of inertia tensor takes the form
\beq
  \langle T_{\mu\nu} \rangle = \left( 
  \begin{array}{@{\,}c|cccc@{\,}}
  {\rm SO(6) \, part} & & & & \\ \hline
  & C & \alpha & \alpha & \alpha \\
  & \alpha & C & \alpha & \alpha \\
  & \alpha & \alpha & C & \alpha \\
  & \alpha & \alpha & \alpha & C
  \end{array}
\right) .
\eeq
This ansatz
is equivalent to 
the ``$\SO(6)$ ansatz'' used in ref.~\cite{Nishimura:2001sx}.

\item $\SO(6)\times\mathbb{Z}_{3(8,9,10)}$

5 parameters ($M,\tilde{M},\tilde{M}',\tilde{m},\tilde{m}'$)
\begin{gather}
  M_\mu = (\underbrace{M,\cdots,M}_6,\tilde{M},\tilde{M}',\tilde{M}',\tilde{M}')
 \ , \nonumber \\
  m_{7,8,9} = m_{7,9,10} = - m_{7,8,10} = \tilde{m} \ , \quad m_{8,9,10} = \tilde{m}'
  \ , \quad \text{and zero otherwise.}
\end{gather}
The moment of inertia tensor takes the form
\beq
  \langle T_{\mu\nu} \rangle = \left( 
  \begin{array}{@{\,}c|c|ccc@{\,}}
  {\rm SO(6) \, part} & & & & \\ \hline
  & C & \beta & \beta & \beta \\ \hline
  & \beta & C' & \alpha & \alpha \\
  & \beta & \alpha & C' & \alpha \\
  & \beta & \alpha & \alpha & C'
  \end{array}
\right) .
\eeq

\item $\SO(6)\times\mathbb{Z}_{2(7,8|1)}$

5 parameters ($M,\tilde{M},\tilde{M}',\tilde{M}'',\tilde{m}$)
\begin{gather}
  M_\mu = (\underbrace{M,\cdots,M}_6,\tilde{M},\tilde{M},
  \tilde{M}',\tilde{M}'') \ , \nonumber \\
  m_{7,9,10} = m_{8,9,10} = \tilde{m} \ , \quad   \text{and zero otherwise.}
\end{gather}
The moment of inertia tensor takes the form
\beq
  \langle T_{\mu\nu} \rangle = \left( 
  \begin{array}{@{\,}c|cc|c|c@{\,}}
  {\rm SO(6) \, part} & & & & \\ \hline
  & C & \alpha & \times & \times \\ 
  & \alpha & C & \times & \times \\ \hline
  & \times & \times & C' & \times \\ \hline
  & \times & \times & \times & C''
  \end{array}
\right) .
\eeq

\end{itemize}


\paragraph{$\SO(5)$ ansatz}


\begin{itemize}

\item $\SO(5)\times\mathbb{Z}_{5(6,7,8,9,10)}$

4 parameters ($M,\tilde{M},\tilde{m},\tilde{m}'$)
\begin{gather}
  M_\mu = (\underbrace{M,\cdots,M}_5,\underbrace{\tilde{M},\cdots,\tilde{M}}_5)
  \ , \nonumber \\  
  m_{6,7,8} = m_{7,8,9} = m_{8,9,10} = m_{6,9,10} = m_{6,7,10} = \tilde{m} \ , \nonumber \\
  m_{6,7,9} = m_{7,8,10} = m_{6,8,9} = m_{7,9,10} = m_{6,8,10} = \tilde{m}'
  \ , \quad \text{and zero otherwise.}
\end{gather}
The moment of inertia tensor takes the form
\beq
  \langle T_{\mu\nu} \rangle = \left( 
  \begin{array}{@{\,}c|ccccc@{\,}}
  {\rm SO(5) \, part} & & & & & \\ \hline
  & C & \alpha & \alpha & \alpha & \alpha \\
  & \alpha & C & \alpha & \alpha & \alpha \\
  & \alpha & \alpha & C & \alpha & \alpha \\
  & \alpha & \alpha & \alpha & C & \alpha \\
  & \alpha & \alpha & \alpha & \alpha & C
  \end{array}
\right) .
\eeq

\item $\SO(5)\times\mathbb{Z}_{4(6,7,8,9|1)}$

5 parameters ($M,\tilde{M},\tilde{M}',\tilde{m},\tilde{m}'$)
\begin{gather}
  M_\mu = (\underbrace{M,\cdots,M}_5,\tilde{M},
    \tilde{M},\tilde{M},\tilde{M},\tilde{M}') \ , \nonumber \\
  m_{6,7,8} = m_{7,8,9} = m_{6,8,9} =  m_{6,7,9} = \tilde{m} \ , \nonumber \\
  m_{6,7,10} = m_{7,8,10} = m_{8,9,10} = -m_{6,9,10} = \tilde{m}' \ , \quad
  \text{and zero otherwise.}
\end{gather}
The moment of inertia tensor takes the form
\beq
  \langle T_{\mu\nu} \rangle = \left( 
  \begin{array}{@{\,}c|cccc|c@{\,}}
  {\rm SO(5) \, part} & & & & & \\ \hline
  & C & \alpha & \alpha & \alpha & \beta \\
  & \alpha & C & \alpha & \alpha & \beta \\
  & \alpha & \alpha & C & \alpha & \beta \\
  & \alpha & \alpha & \alpha & C & \beta \\ \hline
  & \beta & \beta & \beta & \beta & C'
  \end{array}
\right) .
\eeq

\item $\SO(5)\times\mathbb{Z}_{4(6,7,8,9|10)}$

5 parameters ($M,\tilde{M},\tilde{M}',\tilde{m},\tilde{m}'$)
\begin{gather}
  M_\mu = (\underbrace{M,\cdots,M}_5,\tilde{M},\tilde{M},\tilde{M},\tilde{M},
  \tilde{M}') \ , \nonumber \\
  m_{6,7,8} = m_{7,8,9} = m_{6,8,9} = m_{6,7,9} = \tilde{m} \ , \nonumber \\
  m_{6,7,10} = - m_{7,8,10} = m_{8,9,10} = m_{6,9,10} = \tilde{m}' \ , \quad
  \text{and zero otherwise.}
\end{gather}
The moment of inertia tensor takes the form
\beq
  \langle T_{\mu\nu} \rangle = \left( 
  \begin{array}{@{\,}c|cccc|c@{\,}}
  {\rm SO(5) \, part} & & & & & \\ \hline
  & C & \alpha & \alpha & \alpha & 0 \\
  & \alpha & C & \alpha & \alpha & 0 \\
  & \alpha & \alpha & C & \alpha & 0 \\
  & \alpha & \alpha & \alpha & C & 0 \\ \hline
  & 0 & 0 & 0 & 0 & C'
  \end{array}
\right) .
\eeq

\item $\SO(5)\times\mathbb{Z}_{2(6,7|1)}\times\mathbb{Z}_{3(8,9,10)}$

5 parameters ($M,\tilde{M},\tilde{M}',\tilde{m},\tilde{m}'$)
\begin{gather}
  M_\mu = (\underbrace{M,\cdots,M}_5,\tilde{M},\tilde{M},
    \tilde{M}',\tilde{M}',\tilde{M}') \ , \nonumber \\
  m_{6,8,9} = m_{6,9,10} = - m_{6,8,10} = m_{7,8,9} = m_{7,9,10}= - m_{7,8,10} = \tilde{m} \ ,
    \nonumber \\
  m_{8,9,10} = \tilde{m}' \ , \quad
  \text{and zero otherwise.}
\end{gather}
The moment of inertia tensor takes the form
\beq
  \langle T_{\mu\nu} \rangle = \left( 
  \begin{array}{@{\,}c|cc|ccc@{\,}}
  {\rm SO(5) \, part} & & & & & \\ \hline
  & C & \alpha & \beta & \beta & \beta \\
  & \alpha & C & \beta & \beta & \beta \\ \hline
  & \beta & \beta & C' & \gamma & \gamma \\
  & \beta & \beta & \gamma & C' & \gamma \\
  & \beta & \beta & \gamma & \gamma & C'
  \end{array}
\right) .
\eeq

\item $\SO(5)\times\mathbb{Z}_{2(6,7|8,9,10)}\times\mathbb{Z}_{3(8,9,10)}$

5 parameters ($M,\tilde{M},\tilde{M}',\tilde{m},\tilde{m}'$)
\begin{gather}
  M_\mu = (\underbrace{M,\cdots,M}_5,\tilde{M},\tilde{M},
    \tilde{M}',\tilde{M}',\tilde{M}') \ , \nonumber \\
  m_{6,8,9} = m_{7,8,9} = m_{6,9,10} = m_{7,9,10} = - m_{6,8,10} = - m_{7,8,10} = \tilde{m} \ ,
    \nonumber \\
  m_{6,7,8} = m_{6,7,9} = m_{6,7,10} = \tilde{m}' \ , \quad
  \text{and zero otherwise.}
\end{gather}
The moment of inertia tensor takes the form
\beq
  \langle T_{\mu\nu} \rangle = \left( 
  \begin{array}{@{\,}c|cc|ccc@{\,}}
  {\rm SO(5) \, part} & & & & & \\ \hline
  & C & \alpha & 0 & 0 & 0 \\
  & \alpha & C & 0 & 0 & 0 \\ \hline
  & 0 & 0 & C' & \beta & \beta \\
  & 0 & 0 & \beta & C' & \beta \\
  & 0 & 0 & \beta & \beta & C'
  \end{array}
\right) .
\eeq

\item $\SO(5)\times\mathbb{Z}_{2(6,7|1)}\times\mathbb{Z}_{2(8,9|1)}$

5 parameters ($M,\tilde{M},\tilde{M}',\tilde{M}'',\tilde{m}$)
\begin{gather}
  M_\mu = (\underbrace{M,\cdots,M}_5,\tilde{M},\tilde{M},
    \tilde{M}',\tilde{M}',\tilde{M}'') \ , \nonumber \\
  m_{6,8,10} = m_{7,8,10} = m_{6,9,10} = m_{7,9,10}= \tilde{m} \ , \quad
  \text{and zero otherwise.}
\end{gather}
The moment of inertia tensor takes the form
\beq
  \langle T_{\mu\nu} \rangle = \left( 
  \begin{array}{@{\,}c|cc|cc|c@{\,}}
  {\rm SO(5) \, part} & & & & & \\ \hline
  & C & \alpha & \times & \times & \times \\
  & \alpha & C & \times & \times & \times \\ \hline
  & \times & \times & C' & \beta & \times \\
  & \times & \times & \beta & C' & \times \\ \hline
  & \times & \times & \times & \times & C''
  \end{array}
\right) .
\eeq

\item $\SO(5)\times\mathbb{Z}_{2(6,7|10)}\times\mathbb{Z}_{2(8,9|10)}
\times\mathbb{Z}_{2(\{6,7\},\{8,9\}|1,10)}$

5 parameters ($M,\tilde{M},\tilde{M}',\tilde{m},\tilde{m}'$)
\begin{gather}
  M_\mu = (\underbrace{M,\cdots,M}_5,
    \tilde{M},\tilde{M},\tilde{M},\tilde{M},\tilde{M}') \ , \nonumber \\
  m_{6,8,10} = - m_{7,8,10} = - m_{6,9,10} =  m_{7,9,10} = \tilde{m} \ , \nonumber \\
  m_{6,7,10} = - m_{8,9,10} = \tilde{m}' \ , \quad  \text{and zero otherwise.}
\end{gather}
The moment of inertia tensor takes the form
\beq
  \langle T_{\mu\nu} \rangle = \left( 
  \begin{array}{@{\,}c|cc|cc|c@{\,}}
  {\rm SO(5) \, part} & & & & & \\ \hline
  & C & \alpha & \times & \times & 0 \\
  & \alpha & C & \times & \times & 0 \\ \hline
  & \times & \times & C & \alpha & 0 \\
  & \times & \times & \alpha & C & 0 \\ \hline
  & 0 & 0 & 0 & 0 & C'
  \end{array}
\right) .
\eeq

\end{itemize}


\paragraph{$\SO(4)$ ansatz}


\begin{itemize}

\item $\SO(4)\times\mathbb{Z}_{2(5,6|1)}\times\mathbb{Z}_{2(7,8|1)}
\times\mathbb{Z}_{2(9,10|1)}$

5 parameters ($M,\tilde{M},\tilde{M}',\tilde{M}'',\tilde{m}$)
\begin{gather}
  M_\mu = (M,M,M,M,\tilde{M},\tilde{M},\tilde{M}',\tilde{M}',
    \tilde{M}'',\tilde{M}'') \ , \nonumber \\
  m_{5,7,9} = m_{6,7,9} = m_{5,8,9} = m_{6,8,9} =
    m_{5,7,10} = m_{6,7,10} \nonumber \\
  = m_{5,8,10} = m_{6,8,10} = \tilde{m} \ , \quad
  \text{and zero otherwise.}
\end{gather}
The moment of inertia tensor takes the form
\beq
  \langle T_{\mu\nu} \rangle = \left( 
  \begin{array}{@{\,}c|cc|cc|cc@{\,}}
  {\rm SO(4) \, part} & & & & & & \\ \hline
  & C & \alpha & \times & \times & \times & \times \\
  & \alpha & C & \times & \times & \times & \times \\ \hline
  & \times & \times & C' & \beta & \times & \times \\
  & \times & \times & \beta & C' & \times & \times \\ \hline
  & \times & \times & \times & \times & C'' & \gamma \\
  & \times & \times & \times & \times & \gamma & C''
  \end{array}
\right) .
\eeq

By
further imposing 
$\mathbb{Z}_{3(\{5,6\},\{7,8\},\{9,10\})}$,
which enforces
$\tilde{M} = \tilde{M}' = \tilde{M}''$,
one obtains the ``$\SO(4)$ ansatz'' used in ref.~\cite{Nishimura:2001sx}.

\item $\SO(4)\times\mathbb{Z}_{3(5,6,7)}\times\mathbb{Z}_{3(8,9,10)}
\times\mathbb{Z}_{2(\{5,6,7\},\{8,9,10\}|1)}$

4 parameters ($M,\tilde{M},\tilde{m},\tilde{m}'$)
\begin{gather}
  M_\mu = (M,M,M,M,\underbrace{\tilde{M},\cdots,\tilde{M}}_6) \ , \nonumber \\
  m_{5,6,7} = m_{8,9,10} = \tilde{m} \ , \nonumber \\
  m_{5,6,8} = m_{5,6,9} = m_{5,6,10} = m_{6,7,8} = m_{6,7,9} = m_{6,7,10} \nonumber \\
= - m_{5,7,8} = - m_{5,7,9} = - m_{5,7,10} = m_{5,8,9} = m_{6,8,9} = m_{7,8,9} \nonumber \\
 = m_{5,9,10} = m_{6,9,10} = m_{7,9,10}
= - m_{5,8,10} = - m_{6,8,10} = - m_{7,8,10} = \tilde{m}'
 \ , \nonumber \\
  \text{and zero otherwise.}
\end{gather}
The moment of inertia tensor takes the form
\beq
  \langle T_{\mu\nu} \rangle = \left( 
  \begin{array}{@{\,}c|ccc|ccc@{\,}}
  {\rm SO(4) \, part} & & & & & & \\ \hline
  & C & \alpha & \alpha & \beta & \beta & \beta \\
  & \alpha & C & \alpha & \beta & \beta & \beta \\
  & \alpha & \alpha & C & \beta & \beta & \beta \\ \hline
  & \beta & \beta & \beta & C & \alpha & \alpha \\
  & \beta & \beta & \beta & \alpha & C & \alpha \\
  & \beta & \beta & \beta & \alpha & \alpha & C
  \end{array}
\right) .
\eeq

By imposing
$\SO(3)\times\SO(3)$ instead of
$\mathbb{Z}_3\times\mathbb{Z}_3$,
which enforces $\tilde{m}' = 0$,
one obtains the ``$\SO(4)$ ansatz'' used 
in refs.~\cite{%
Kawai:2002jk,Kawai:2002ub,Aoyama:2006di,Aoyama:2006rk,Aoyama:2006je}.

\item $\SO(4)\times\mathbb{Z}_{4(5,6,7,8|1)}\times\mathbb{Z}_{2(9,10|1)}$

5 parameters ($M,\tilde{M},\tilde{M}',\tilde{m},\tilde{m}'$)
\begin{gather}
  M_\mu = (M,M,M,M,\tilde{M},\tilde{M},\tilde{M},\tilde{M},
    \tilde{M}',\tilde{M}') \ , \nonumber \\
  m_{5,6,9} = m_{6,7,9} = m_{7,8,9} = - m_{5,8,9}
    = m_{5,6,10} = m_{6,7,10} = m_{7,8,10} = - m_{5,8,10}= \tilde{m} \ , \nonumber \\
  m_{5,6,7} = m_{6,7,8} = m_{5,7,8} = m_{5,6,8} = \tilde{m}' \ , \quad
  \text{and zero otherwise.}
\end{gather}
The moment of inertia tensor takes the form
\beq
  \langle T_{\mu\nu} \rangle = \left( 
  \begin{array}{@{\,}c|cccc|cc@{\,}}
  {\rm SO(4) \, part} & & & & & & \\ \hline
  & C & \alpha & \alpha & \alpha & \beta & \beta \\
  & \alpha & C & \alpha & \alpha & \beta & \beta \\
  & \alpha & \alpha & C & \alpha & \beta & \beta \\
  & \alpha & \alpha & \alpha & C & \beta & \beta \\ \hline
  & \beta & \beta & \beta & \beta & C' & \gamma \\
  & \beta & \beta & \beta & \beta & \gamma & C'
  \end{array}
\right) .
\eeq

\item $\SO(4)\times\mathbb{Z}_{5(6,7,8,9,10)}\times\mathbb{Z}_{2(-|1,6,7,8,9,10)}$

5 parameters ($M,\tilde{M},\tilde{M}',\tilde{m},\tilde{m}'$)
\begin{gather}
  M_\mu = (M,M,M,M,\tilde{M},\underbrace{\tilde{M}',\cdots,\tilde{M}'}_5)
    \ , \nonumber \\
  m_{5,6,7} = m_{5,7,8} = m_{5,8,9} = m_{5,9,10} = - m_{5,6,10} = \tilde{m} \ , \nonumber \\
  m_{5,6,8} = m_{5,7,9} = m_{5,8,10} = - m_{5,6,9} = - m_{5,7,10} = \tilde{m}' \ , \quad
  \text{and zero otherwise.}
\end{gather}
The moment of inertia tensor takes the form
\beq
  \langle T_{\mu\nu} \rangle = \left( 
  \begin{array}{@{\,}c|c|ccccc@{\,}}
  {\rm SO(4) \, part} & & & & & & \\ \hline
  & C & \times & \times & \times & \times & \times \\ \hline
  & \times & C' & \alpha & \alpha & \alpha & \alpha \\
  & \times & \alpha & C' & \alpha & \alpha & \alpha \\
  & \times & \alpha & \alpha & C' & \alpha & \alpha \\
  & \times & \alpha & \alpha & \alpha & C' & \alpha \\
  & \times & \alpha & \alpha & \alpha & \alpha & C'
  \end{array}
\right) .
\eeq

\item $\SO(4)\times\mathbb{Z}_{3(5,6,7)}\times\mathbb{Z}_{3(8,9,10)}
\times\mathbb{Z}_{2(-|1,8,9,10)}$

5 parameters ($M,\tilde{M},\tilde{M}',\tilde{m},\tilde{m}'$)
\begin{gather}
  M_\mu = (M,M,M,M,\tilde{M},\tilde{M},\tilde{M},
    \tilde{M}',\tilde{M}',\tilde{M}') \ , \nonumber \\
  m_{5,8,9} = m_{5,9,10} = - m_{5,8,10} = m_{6,8,9} = m_{6,9,10} 
= - m_{6,8,10} \nonumber \\
  = m_{7,8,9} = m_{7,9,10} = - m_{7,8,10} = \tilde{m} \ , \nonumber \\
  m_{5,6,7} = \tilde{m}' \ , \quad
  \text{and zero otherwise.}
\end{gather}
The moment of inertia tensor takes the form
\beq
  \langle T_{\mu\nu} \rangle = \left( 
  \begin{array}{@{\,}c|ccc|ccc@{\,}}
  {\rm SO(4) \, part} & & & & & & \\ \hline
  & C & \alpha & \alpha & 0 & 0 & 0 \\
  & \alpha & C & \alpha & 0 & 0 & 0 \\
  & \alpha & \alpha & C & 0 & 0 & 0 \\ \hline
  & 0 & 0 & 0 & C' & \beta & \beta \\
  & 0 & 0 & 0 & \beta & C' & \beta \\
  & 0 & 0 & 0 & \beta & \beta & C'
  \end{array}
\right) .
\eeq

\end{itemize}


\paragraph{$\SO(3)$ ansatz}


\begin{itemize}

\item $\SO(3)\times\mathbb{Z}_{2(4,5|1)}\times\mathbb{Z}_{2(6,7|1)}
\times\mathbb{Z}_{2(8,9|1)}\times\mathbb{Z}_{3(\{4,5\},\{6,7\},\{8,9\})}$

5 parameters ($M,\tilde{M},\tilde{M}',\tilde{m},\tilde{m}'$)
\begin{gather}
  M_\mu = (M,M,M,\underbrace{\tilde{M},\cdots,\tilde{M}}_6,\tilde{M}') \ ,
    \nonumber \\
  m_{4,6,8} = m_{5,6,8} = m_{4,7,8} = m_{5,7,8} =
    m_{4,6,9} = m_{5,6,9} = m_{4,7,9} = m_{5,7,9} = \tilde{m} \ , \nonumber \\
  m_{4,6,10} = m_{5,6,10} = m_{4,7,10} = m_{5,7,10} = - m_{4,8,10} = - m_{5,8,10} \nonumber \\
= - m_{4,9,10} = - m_{5,9,10} =
    m_{6,8,10} = m_{7,8,10} = m_{6,9,10} = m_{7,9,10} = \tilde{m}' \ , \nonumber \\
  \text{and zero otherwise.}
\end{gather}
The moment of inertia tensor takes the form
\beq
  \langle T_{\mu\nu} \rangle = \left( 
  \begin{array}{@{\,}c|cc|cc|cc|c@{\,}}
  {\rm SO(3) \, part} & & &  & & & & \\ \hline
  & C & \alpha & \beta & \beta & \beta & \beta & \gamma \\
  & \alpha & C & \beta & \beta & \beta & \beta & \gamma \\ \hline
  & \beta & \beta & C & \alpha & \beta & \beta & \gamma \\
  & \beta & \beta & \alpha & C & \beta & \beta & \gamma \\ \hline
  & \beta & \beta & \beta & \beta & C & \alpha & \gamma \\
  & \beta & \beta & \beta & \beta & \alpha & C & \gamma \\ \hline
  & \gamma & \gamma & \gamma & \gamma & \gamma & \gamma & C'
  \end{array}
\right) .
\eeq

\item $\SO(3)\times\mathbb{Z}_{2(4,5|10)}\times\mathbb{Z}_{2(6,7|10)}
\times\mathbb{Z}_{2(8,9|10)}\times\mathbb{Z}_{3(\{4,5\},\{6,7\},\{8,9\})}
\times\mathbb{Z}_{2(\{4,5\},\{6,7\})}$

5 parameters ($M,\tilde{M},\tilde{M}',m,\tilde{m}$)
\begin{gather}
  M_\mu = (M,M,M,\underbrace{\tilde{M},\cdots,\tilde{M}}_6,\tilde{M}') \ ,
    \nonumber \\
  m_{1,2,3} = m \ , \quad
  m_{4,5,10} = m_{6,7,10} = m_{8,9,10} = \tilde{m} \ , \quad
  \text{and zero otherwise.}
\end{gather}
The moment of inertia tensor takes the form
\beq
  \langle T_{\mu\nu} \rangle = \left( 
  \begin{array}{@{\,}c|cc|cc|cc|c@{\,}}
  {\rm SO(3) \, part} & & &  & & & & \\ \hline
  & C & \times & \times & \times & \times & \times & 0 \\
  & \times & C & \times & \times & \times & \times & 0 \\ \hline
  & \times & \times & C & \times & \times & \times & 0 \\
  & \times & \times & \times & C & \times & \times & 0 \\ \hline
  & \times & \times & \times & \times & C & \times & 0 \\
  & \times & \times & \times & \times & \times & C & 0 \\ \hline
  & 0 & 0 & 0 & 0 & 0 & 0 & C'
  \end{array}
\right) .
\eeq

\item $\SO(3)\times\mathbb{Z}_{2(4,5|10)}\times\mathbb{Z}_{2(6,7|10)}
\times\mathbb{Z}_{2(8,9|10)}\times\mathbb{Z}_{3(\{4,5\},\{6,7\},\{8,9\})}
\times\mathbb{Z}_{2(-|1,10)}$

4 parameters ($M,\tilde{M},\tilde{M}',\tilde{m}$)
\begin{gather}
  M_\mu = (M,M,M,\underbrace{\tilde{M},\cdots,\tilde{M}}_6,\tilde{M}') \ ,
    \nonumber \\
  m_{4,6,8} = m_{5,6,8} = m_{4,7,8} = m_{5,7,8} =
    m_{4,6,9} = m_{5,6,9} = m_{4,7,9} = m_{5,7,9} = \tilde{m} \ , \nonumber \\
  \text{and zero otherwise.}
\end{gather}
The moment of inertia tensor takes the form
\beq
  \langle T_{\mu\nu} \rangle = \left( 
  \begin{array}{@{\,}c|cc|cc|cc|c@{\,}}
  {\rm SO(3) \, part} & & &  & & & & \\ \hline
  & C & \alpha & \times & \times & \times & \times & 0 \\
  & \alpha & C & \times & \times & \times & \times & 0 \\ \hline
  & \times & \times & C & \alpha & \times & \times & 0 \\
  & \times & \times & \alpha & C & \times & \times & 0 \\ \hline
  & \times & \times & \times & \times & C & \alpha & 0 \\
  & \times & \times & \times & \times & \alpha & C & 0 \\ \hline
  & 0 & 0 & 0 & 0 & 0 & 0 & C'
  \end{array}
\right) .
\eeq

\item $\SO(3)\times\mathbb{Z}_{3(4,5,6)}\times\mathbb{Z}_{3(7,8,9)}
\times\mathbb{Z}_{2(\{4,5,6\},\{7,8,9\}|1)}\times\mathbb{Z}_{2(-|1,10)}$

5 parameters ($M,\tilde{M},\tilde{M}',\tilde{m},\tilde{m}'$)
\begin{gather}
  M_\mu = (M,M,M,\underbrace{\tilde{M},\cdots,\tilde{M}}_6,\tilde{M}') \ ,
    \nonumber \\
  m_{4,5,\nu} = m_{5,6,\nu} = - m_{4,6,\nu}
 = m_{7,8,\mu} = m_{8,9,\mu} = - m_{7,9,\mu} = \tilde{m} \ , \nonumber \\
\mbox{where~$\mu = 4,5,6$ and $\nu = 7,8,9$} \ , \nonumber \\
  m_{4,5,6} = m_{7,8,9} = \tilde{m}' \ , \quad
  \text{and zero otherwise.}
\end{gather}
The moment of inertia tensor takes the form
\beq
  \langle T_{\mu\nu} \rangle = \left( 
  \begin{array}{@{\,}c|ccc|ccc|c@{\,}}
  {\rm SO(3) \, part} & & & & & & & \\ \hline
  & C & \alpha & \alpha & \beta & \beta & \beta & 0 \\
  & \alpha & C & \alpha & \beta & \beta & \beta & 0 \\ 
  & \alpha & \alpha & C & \beta & \beta & \beta & 0 \\ \hline
  & \beta & \beta & \beta & C & \alpha & \alpha & 0 \\
  & \beta & \beta & \beta & \alpha & C & \alpha & 0 \\
  & \beta & \beta & \beta & \alpha & \alpha & C & 0 \\ \hline
  & 0 & 0 & 0 & 0 & 0 & 0 & C'
  \end{array}
\right) .
\eeq

\item $\SO(3)\times\mathbb{Z}_{3(4,5,6)}\times\mathbb{Z}_{2(7,8|1)}
\times\mathbb{Z}_{2(9,10|1)}\times\mathbb{Z}_{2(\{7,8\},\{9,10\})}$

5 parameters ($M,\tilde{M},\tilde{M}',\tilde{m},\tilde{m}'$)
\begin{gather}
  M_\mu = (M,M,M,\tilde{M},\tilde{M},\tilde{M},
    \tilde{M}',\tilde{M}',\tilde{M}',\tilde{M}') \ , \nonumber \\
  m_{4,5,7} = m_{4,5,8} = m_{5,6,7} = m_{5,6,8} = - m_{4,6,7} = - m_{4,6,8} \nonumber \\
  = m_{4,5,9} = m_{4,5,10} = m_{5,6,9} = m_{5,6,10} = - m_{4,6,9} = - m_{4,6,10} 
  = \tilde{m} \ , \nonumber \\
  m_{4,5,6} = \tilde{m}' \ , \quad
  \text{and zero otherwise.}
\end{gather}
The moment of inertia tensor takes the form
\beq
  \langle T_{\mu\nu} \rangle = \left( 
  \begin{array}{@{\,}c|ccc|cc|cc@{\,}}
  {\rm SO(3) \, part} & & & & & & & \\ \hline
  & C & \alpha & \alpha & \gamma & \gamma & \gamma & \gamma \\
  & \alpha & C & \alpha & \gamma & \gamma & \gamma & \gamma \\
  & \alpha & \alpha & C & \gamma & \gamma & \gamma & \gamma \\ \hline
  & \gamma & \gamma & \gamma & C' & \beta & \beta & \beta \\
  & \gamma & \gamma & \gamma & \beta & C' & \beta & \beta \\ \hline
  & \gamma & \gamma & \gamma & \beta & \beta & C' & \beta \\
  & \gamma & \gamma & \gamma & \beta & \beta & \beta & C'
  \end{array}
\right) .
\eeq

\item $\SO(3)\times\mathbb{Z}_{3(4,5,6)}\times\mathbb{Z}_{4(7,8,9,10|1)}
\times\mathbb{Z}_{2(-|7,8,9,10)}$

5 parameters ($M,\tilde{M},\tilde{M}',\tilde{m},\tilde{m}'$)
\begin{gather}
  M_\mu = (M,M,M,\tilde{M},\tilde{M},\tilde{M},
    \tilde{M}',\tilde{M}',\tilde{M}',\tilde{M}') \ , \nonumber \\
  m_{4,7,8} = m_{4,8,9} = m_{4,9,10} = - m_{4,7,10} = m_{5,7,8} = m_{5,8,9} \nonumber \\
  = m_{5,9,10} = - m_{5,7,10} = m_{6,7,8} = m_{6,8,9} = m_{6,9,10} = - m_{6,7,10} 
  = \tilde{m} \ , \nonumber \\
  m_{4,5,6} = \tilde{m}' \ , \quad \text{and zero otherwise.}
\end{gather}
The moment of inertia tensor takes the form
\beq
  \langle T_{\mu\nu} \rangle = \left( 
  \begin{array}{@{\,}c|ccc|cccc@{\,}}
  {\rm SO(3) \, part} & & & & & & & \\ \hline
  & C & \alpha & \alpha & 0 & 0 & 0 & 0 \\
  & \alpha & C & \alpha & 0 & 0 & 0 & 0 \\
  & \alpha & \alpha & C & 0 & 0 & 0 & 0 \\ \hline
  & 0 & 0 & 0 & C' & \times & \times & \times \\
  & 0 & 0 & 0 & \times & C' & \times & \times \\
  & 0 & 0 & 0 & \times & \times & C' & \times \\
  & 0 & 0 & 0 & \times & \times & \times & C'
  \end{array}
\right) .
\eeq

\item $\SO(3)\times\mathbb{Z}_{2(4,5|1)}\times\mathbb{Z}_{5(6,7,8,9,10)}
\times\mathbb{Z}_{2(-|1,6,7,8,9,10)}$

5 parameters ($M,\tilde{M},\tilde{M}',\tilde{m},\tilde{m}'$)
\begin{gather}
  M_\mu = (M,M,M,\tilde{M},\tilde{M},\underbrace{\tilde{M}',\cdots,\tilde{M}'}_5)
    \ , \nonumber \\
  m_{4,6,7} = m_{4,7,8} = m_{4,8,9} = m_{4,9,10} = - m_{4,6,10} \nonumber \\
  = m_{5,6,7} = m_{5,7,8} = m_{5,8,9} = m_{5,9,10} = - m_{5,6,10} = \tilde{m} \ , 
    \nonumber \\
  m_{4,6,8} = m_{4,7,9} = m_{4,8,10} = - m_{4,6,9} = - m_{4,7,10} \nonumber \\
  = m_{5,6,8} = m_{5,7,9} = m_{5,8,10} = - m_{5,6,9} = - m_{5,7,10} = \tilde{m}' \ ,
    \nonumber \\
  \text{and zero otherwise.}
\end{gather}
The moment of inertia tensor takes the form
\beq
  \langle T_{\mu\nu} \rangle = \left( 
  \begin{array}{@{\,}c|cc|ccccc@{\,}}
  {\rm SO(3) \, part} & & & & & & & \\ \hline
  & C & \alpha & 0 & 0 & 0 & 0 & 0 \\
  & \alpha & C & 0 & 0 & 0 & 0 & 0 \\ \hline
  & 0 & 0 & C' & \beta & \beta & \beta & \beta \\
  & 0 & 0 & \beta & C' & \beta & \beta & \beta \\
  & 0 & 0 & \beta & \beta & C' & \beta & \beta \\
  & 0 & 0 & \beta & \beta & \beta & C' & \beta \\
  & 0 & 0 & \beta & \beta & \beta & \beta & C'
  \end{array}
\right) .
\eeq

\item $\SO(3)\times\mathbb{Z}_{6(5,6,7,8,9,10|1)}\times\mathbb{Z}_{2(-|5,6,7,8,9,10)}$

5 parameters ($M,\tilde{M},\tilde{M}',\tilde{m},\tilde{m}'$)
\begin{gather}
  M_\mu = (M,M,M,\tilde{M},\underbrace{\tilde{M}',\cdots,\tilde{M}'}_6) \ ,
    \nonumber \\
  m_{4,5,6} = m_{4,6,7} = m_{4,7,8} = m_{4,8,9} = m_{4,9,10} = - m_{4,5,10} = \tilde{m} \ ,
    \nonumber \\
  m_{4,5,7} = m_{4,6,8} = m_{4,7,9} = m_{4,8,10} = - m_{4,5,9} = - m_{4,6,10} = \tilde{m}' \ ,
    \nonumber \\
  \text{and zero otherwise.}
\end{gather}
The moment of inertia tensor takes the form
\beq
  \langle T_{\mu\nu} \rangle = \left( 
  \begin{array}{@{\,}c|c|cccccc@{\,}}
  {\rm SO(3) \, part} & & & & & & & \\ \hline
  & C & 0 & 0 & 0 & 0 & 0 & 0 \\ \hline
  & 0 & C' & \alpha & \alpha & \alpha & \alpha & \alpha \\
  & 0 & \alpha & C' & \alpha & \alpha & \alpha & \alpha \\
  & 0 & \alpha & \alpha & C' & \alpha & \alpha & \alpha \\
  & 0 & \alpha & \alpha & \alpha & C' & \alpha & \alpha \\
  & 0 & \alpha & \alpha & \alpha & \alpha & C' & \alpha \\
  & 0 & \alpha & \alpha & \alpha & \alpha & \alpha & C'
  \end{array}
\right) .
\eeq

\item $\SO(3)\times\mathbb{Z}_{2(4,5|1)}\times\mathbb{Z}_{5(6,7,8,9,10|4,5)}$

5 parameters ($M,\tilde{M},\tilde{M}',\tilde{m},\tilde{m}'$)
\begin{gather}
  M_\mu = (M,M,M,\tilde{M},\tilde{M},
    \underbrace{\tilde{M}',\cdots,\tilde{M}'}_5) \ ,
    \nonumber \\
  m_{6,7,8} = m_{7,8,9} = m_{8,9,10} = m_{6,9,10} = m_{6,7,10} = \tilde{m} \ ,
    \nonumber \\
  m_{6,7,9} = m_{7,8,10} = m_{6,8,9} = m_{7,9,10} = m_{6,8,10} = \tilde{m}' \ ,
    \nonumber \\
  \text{and zero otherwise.}
\end{gather}
The moment of inertia tensor takes the form
\beq
  \langle T_{\mu\nu} \rangle = \left( 
  \begin{array}{@{\,}c|cc|ccccc@{\,}}
  {\rm SO(3) \, part} & & & & & & & \\ \hline
  & C & \times & 0 & 0 & 0 & 0 & 0 \\
  & \times & C & 0 & 0 & 0 & 0 & 0 \\ \hline
  & 0 & 0 & C' & \alpha & \alpha & \alpha & \alpha \\
  & 0 & 0 & \alpha & C' & \alpha & \alpha & \alpha \\
  & 0 & 0 & \alpha & \alpha & C' & \alpha & \alpha \\
  & 0 & 0 & \alpha & \alpha & \alpha & C' & \alpha \\
  & 0 & 0 & \alpha & \alpha & \alpha & \alpha & C'
  \end{array}
\right) .
\eeq

\end{itemize}


\paragraph{$\SO(2)$ ansatz}


\begin{itemize}

\item $\SO(2)\times\mathbb{Z}_{2(3,4|1)}\times\mathbb{Z}_{2(5,6|1)}
\times\mathbb{Z}_{2(7,8|1)}\times\mathbb{Z}_{2(9,10|1)}
\times\mathbb{Z}_{4(\{3,4\},\{5,6\},\{7,8\},\{9,10\})}$

3 parameters ($M,\tilde{M},\tilde{m}$)
\begin{gather}
  M_\mu = (M,M,\underbrace{\tilde{M},\cdots,\tilde{M}}_8) \ , \nonumber \\
  m_{3,5,7} = m_{3,6,7} = m_{3,5,8} = m_{3,6,8}
 = m_{4,5,7} = m_{4,6,7} = m_{4,5,8} = m_{4,6,8} \nonumber \\
 = m_{5,7,9} = m_{5,8,9} = m_{5,7,10} = m_{5,8,10}
 = m_{6,7,9} = m_{6,8,9} = m_{6,7,10} = m_{6,8,10} \nonumber \\
 = m_{3,7,9} = m_{3,8,9} = m_{3,7,10} = m_{3,8,10}
 = m_{4,7,9} = m_{4,8,9} = m_{4,7,10} = m_{4,8,10} \nonumber \\
 = m_{3,5,9} = m_{3,6,9} = m_{3,5,10} = m_{3,6,10}
 = m_{4,5,9} = m_{4,6,9} = m_{4,5,10} = m_{4,6,10} = \tilde{m} \ , \nonumber \\
  \text{and zero otherwise.}
\end{gather}
The moment of inertia tensor takes the form
\beq
  \langle T_{\mu\nu} \rangle = \left( 
  \begin{array}{@{\,}c|cc|cc|cc|cc@{\,}}
  {\rm SO(2) \, part} & & & & & & & \\ \hline
  & C & \alpha & \beta & \beta & \beta & \beta & \beta & \beta \\
  & \alpha & C & \beta & \beta & \beta & \beta & \beta & \beta \\ \hline
  & \beta & \beta & C & \alpha & \beta & \beta & \beta & \beta \\
  & \beta & \beta & \alpha & C & \beta & \beta & \beta & \beta \\ \hline
  & \beta & \beta & \beta & \beta & C & \alpha & \beta & \beta \\
  & \beta & \beta & \beta & \beta & \alpha & C & \beta & \beta \\ \hline
  & \beta & \beta & \beta & \beta & \beta & \beta & C & \alpha \\
  & \beta & \beta & \beta & \beta & \beta & \beta & \alpha & C
  \end{array}
\right) .
\eeq
This ansatz 
is equivalent to 
the ``SO(2) ansatz'' used in ref.~\cite{Nishimura:2001sx}.

\item $\SO(2)\times\mathbb{Z}_{4(3,4,5,6|1)}\times\mathbb{Z}_{4(7,8,9,10|1)}
\times\mathbb{Z}_{2(\{3,4,5,6\},\{7,8,9,10\})}$

4 parameters ($M,\tilde{M},\tilde{m},\tilde{m}'$)
\begin{gather}
  M_\mu = (M,M,\underbrace{\tilde{M},\cdots,\tilde{M}}_8) \ , \nonumber \\
  m_{3,4,5} = m_{4,5,6} = m_{3,5,6} = m_{3,4,6}
 = m_{7,8,9} = m_{8,9,10} = m_{7,9,10} = m_{7,8,10} = \tilde{m} \ , \nonumber \\
  m_{3,4,7} = m_{4,5,7} = m_{5,6,7} = - m_{3,6,7}
 = m_{3,4,8} = m_{4,5,8} = m_{5,6,8} = - m_{3,6,8} \nonumber \\
 = m_{3,4,9} = m_{4,5,9} = m_{5,6,9} = - m_{3,6,9}
 = m_{3,4,10} = m_{4,5,10} = m_{5,6,10} = - m_{3,6,10} \nonumber \\
 = m_{3,7,8} = m_{3,8,9} = m_{3,9,10} = - m_{3,7,10}
 = m_{4,7,8} = m_{4,8,9} = m_{4,9,10} = - m_{4,7,10} \nonumber \\
 = m_{5,7,8} = m_{5,8,9} = m_{5,9,10} = - m_{5,7,10}
 = m_{6,7,8} = m_{6,8,9} = m_{6,9,10} = - m_{6,7,10} = \tilde{m}' \ , \nonumber \\
  \text{and zero otherwise.}
\end{gather}
The moment of inertia tensor takes the form
\beq
  \langle T_{\mu\nu} \rangle = \left( 
  \begin{array}{@{\,}c|cccc|cccc@{\,}}
  {\rm SO(2) \, part} & & & & & & & \\ \hline
  & C & \alpha & \alpha & \alpha & \beta & \beta & \beta & \beta \\
  & \alpha & C & \alpha & \alpha & \beta & \beta & \beta & \beta \\ 
  & \alpha & \alpha & C & \alpha & \beta & \beta & \beta & \beta \\
  & \alpha & \alpha & \alpha & C & \beta & \beta & \beta & \beta \\ \hline
  & \beta & \beta & \beta & \beta & C & \alpha & \alpha & \alpha \\
  & \beta & \beta & \beta & \beta & \alpha & C & \alpha & \alpha \\
  & \beta & \beta & \beta & \beta & \alpha & \alpha & C & \alpha \\
  & \beta & \beta & \beta & \beta & \alpha & \alpha & \alpha & C
  \end{array}
\right) .
\eeq

\item $\SO(2)\times\mathbb{Z}_{3(3,4,5)}\times\mathbb{Z}_{3(6,7,8)}
\times\mathbb{Z}_{2(\{3,4,5\},\{6,7,8\}|1)}\times\mathbb{Z}_{2(9,10|1)}
\times\mathbb{Z}_{3(-|3,4,5,6,7,8)}$

5 parameters ($M,\tilde{M},\tilde{M}',\tilde{m},\tilde{m}'$)
\begin{gather}
  M_\mu = (M,M,\underbrace{\tilde{M},\cdots,\tilde{M}}_6
    ,\tilde{M}',\tilde{M}') \ , \nonumber \\
  m_{3,4,9} = m_{4,5,9} = - m_{3,5,9} = m_{6,7,9} = m_{7,8,9} = - m_{6,8,9} \nonumber \\
 =  m_{3,4,10} = m_{4,5,10} = - m_{3,5,10} = m_{6,7,10} = m_{7,8,10} = - m_{6,8,10} = \tilde{m}
    \ , \nonumber \\
  m_{3,6,9} = m_{4,6,9} = m_{5,6,9} = m_{3,7,9} = m_{4,7,9} = m_{5,7,9} \nonumber \\
 = m_{3,8,9} = m_{4,8,9} = m_{5,8,9} = m_{3,6,10} = m_{4,6,10} = m_{5,6,10} \nonumber \\
 = m_{3,7,10} = m_{4,7,10} = m_{5,7,10}
 = m_{3,8,10} = m_{4,8,10} = m_{5,8,10} = \tilde{m}' \ , \nonumber \\
  \text{and zero otherwise.}
\end{gather}
The moment of inertia tensor takes the form
\beq
  \langle T_{\mu\nu} \rangle = \left( 
  \begin{array}{@{\,}c|ccc|ccc|cc@{\,}}
  {\rm SO(2) \, part} & & & & & & & \\ \hline
  & C & \alpha & \alpha & \times & \times & \times & 0 & 0 \\
  & \alpha & C & \alpha & \times & \times & \times & 0 & 0 \\ 
  & \alpha & \alpha & C & \times & \times & \times & 0 & 0 \\ \hline
  & \times & \times & \times & C & \alpha & \alpha & 0 & 0 \\ 
  & \times & \times & \times & \alpha & C & \alpha & 0 & 0 \\
  & \times & \times & \times & \alpha & \alpha & C & 0 & 0 \\ \hline
  & 0 & 0 & 0 & 0 & 0 & 0 & C' & \beta \\
  & 0 & 0 & 0 & 0 & 0 & 0 & \beta & C'
  \end{array}
\right) .
\eeq

\item $\SO(2)\times\mathbb{Z}_{3(3,4,5)}\times\mathbb{Z}_{3(6,7,8)}
\times\mathbb{Z}_{2(\{3,4,5\},\{6,7,8\}|1)}\times\mathbb{Z}_{2(9,10|1)}
\times\mathbb{Z}_{2(-|9,10)}$

5 parameters ($M,\tilde{M},\tilde{M}',\tilde{m},\tilde{m}'$)
\begin{gather}
  M_\mu = (M,M,\underbrace{\tilde{M},\cdots,\tilde{M}}_6,
    \tilde{M}',\tilde{M}') \ , \nonumber \\
  m_{3,4,6} = m_{4,5,6} = - m_{3,5,6} = m_{3,4,7} = m_{4,5,7} = - m_{3,5,7} \nonumber \\
 = m_{3,4,8} = m_{4,5,8} = - m_{3,5,8} = m_{3,6,7} = m_{3,7,8} = - m_{3,6,8} \nonumber \\
 = m_{4,6,7} = m_{4,7,8} = - m_{4,6,8}
 = m_{5,6,7} = m_{5,7,8} = - m_{5,6,8} = \tilde{m} \ , \nonumber \\
  m_{3,4,5} = m_{6,7,8} = \tilde{m}' \ , \quad \text{and zero otherwise.}
\end{gather}
The moment of inertia tensor takes the form
\beq
  \langle T_{\mu\nu} \rangle = \left( 
  \begin{array}{@{\,}c|ccc|ccc|cc@{\,}}
  {\rm SO(2) \, part} & & & & & & & \\ \hline
  & C & \alpha & \alpha & \beta & \beta & \beta & 0 & 0 \\
  & \alpha & C & \alpha & \beta & \beta & \beta & 0 & 0 \\ 
  & \alpha & \alpha & C & \beta & \beta & \beta & 0 & 0 \\ \hline
  & \beta & \beta & \beta & C & \alpha & \alpha & 0 & 0 \\ 
  & \beta & \beta & \beta & \alpha & C & \alpha & 0 & 0 \\
  & \beta & \beta & \beta & \alpha & \alpha & C & 0 & 0 \\ \hline
  & 0 & 0 & 0 & 0 & 0 & 0 & C' & \times \\
  & 0 & 0 & 0 & 0 & 0 & 0 & \times & C'
  \end{array}
\right) .
\eeq

\end{itemize}

\section{Free energy from the Krauth-Nicolai-Staudacher conjecture} 
\label{sec:KNSresult}

In this section we derive the value
of free energy from
the analytic formula for the partition function
conjectured by Krauth, Nicolai and Staudacher (KNS) \cite{Krauth:1998xh}.
This conjecture was obtained by
combining 
earlier analytic works \cite{Green:1997tn,Moore:1998et}
with Monte Carlo results at small $N$ \cite{Krauth:1998xh}.
For the $D=10$ model, the formula reads
\begin{eqnarray}
  Z_{\rm KNS} &=& \int \dd A \, \dd \Psi 
\, e^{- S_{\rm KNS}} 
  =  \dfrac{2^{\frac{N(N+1)}{2}} \pi^{\frac{N-1}{2}}}
	{2\sqrt{N}\prod_{k=1}^{N-1} k!}
  \times
  \sum_{m|N} \frac{1}{m^2} \ ,
  \label{eq:analytic-kns} \\
  S_{\rm KNS} &=& \frac{2}{N}(S_{\rm b} + S_{\rm f}) \ ,
\label{action-KNS}
\end{eqnarray}
where $S_{\rm b}$ and 
$S_{\rm f}$ are defined by (\ref{eq:sb}) and (\ref{eq:sf}) respectively,
and the sum runs over all the divisors of $N$. 
The value of this sum is smaller than
$\sum_{m=1}^\infty \frac{1}{m^2} = \frac{\pi^2}{6}$
for arbitrary $N$,
and hence it does not contribute to
the ``free energy density''
(\ref{eq:fedensity})
in the large-$N$ limit.

As one can see from (\ref{action-KNS}),
the definition of the action 
$S_{\rm KNS}$
differs from ours
by the factor of $2/N$. 
In order to absorb this factor, we introduce the rescaled variables 
$A'_\mu = (2/N)^{1/4} A_\mu$ and
$\Psi'_\alpha = (2/N)^{3/8} \Psi_\alpha$,
whose integration measure 
is given by
\begin{equation}
\dd A' \, \dd \Psi' =
 \left( \frac{N}{2} \right)^{\frac{7}{2}(N^2-1)} 
\dd A \, \dd \Psi \ .
\end{equation}
As a result, the partition function
(\ref{eq:10dpf}) can be obtained as
\beq
  Z = \left(\frac{N}{2}\right)^{\frac{7}{2}(N^2 - 1)} 
Z_{\rm KNS} 
  = 2^{-3 N^2 +\frac{N}{2} 
+\frac{5}{2}} \pi^{\frac{N-1}{2}} N^{\frac{7}{2}N^2 - 4} 
\left( \prod_{k=1}^{N-1} k! \right)^{-1} \times \sum_{m|N} \frac{1}{m^2} \  .
\eeq
{}From this, we obtain the large-$N$ asymptotics
for $F = - \log Z$ as
\begin{equation}
  \frac{F}{N^2 - 1} = 
  - 3 \log N + \left( \log 8 - \frac{3}{4} \right)
  + O\left( \frac{\log N}{N^2} \right)       \ .
  \label{eq:analytic}
\end{equation}
The Gaussian expansion method
reproduces the first term of (\ref{eq:analytic})
correctly 
for any ansatz.
Substituting (\ref{eq:analytic})
into the definition (\ref{eq:fedensity}) of 
the free energy density,
we obtain $f = \log 8 -\frac{3}{4}=1.32944...$
as a prediction from the KNS conjecture.






\begin{thebibliography}{99}

\bibitem{Banks:1996vh}
  T.~Banks, W.~Fischler, S.~H.~Shenker and L.~Susskind,
\emph{M theory as a matrix model: A conjecture},
\emph{Phys. Rev.}  {\bf D~ 55} (1997) 5112
[{\tt hep-th/9610043}].


\bibitem{Ishibashi:1996xs}
  N.~Ishibashi, H.~Kawai, Y.~Kitazawa and A.~Tsuchiya,
\emph{A large N reduced model as superstring},
\emph{Nucl.\ Phys.} {\bf B~498} (1997) 467
[{\tt hep-th/9612115}].

\bibitem{Dijkgraaf:1997vv}
  R.~Dijkgraaf, E.~P.~Verlinde and H.~L.~Verlinde,
\emph{Matrix string theory},
\emph{Nucl.\ Phys.} {\bf B~500} (1997) 43
[{\tt hep-th/9703030}].

\bibitem{Aoki:1998vn}
  H.~Aoki, S.~Iso, H.~Kawai, Y.~Kitazawa and T.~Tada,
\emph{Space-time structures from IIB matrix model},
\ptp{99}{1998}{713}
[\hepth{9802085}].


\bibitem{ncym}
%
  H.~Steinacker,
\emph{Emergent geometry and gravity from matrix models: An introduction},
\emph{Class.\ Quant.\ Grav.}  {\bf 27}, 133001 (2010);\\
%
  J.~Lee and H.~S.~Yang,
\emph{Quantum gravity from noncommutative spacetime},
{\tt arXiv:1004.0745}.


\bibitem{Nishimura:2001sx}
  J.~Nishimura and F.~Sugino,
\emph{Dynamical generation of four-dimensional space-time 
in the IIB matrix model},
\emph{JHEP} {\bf 05} (2002) 001 
[{\tt hep-th/0111102}].




\bibitem{Kawai:2002jk}
  H.~Kawai, S.~Kawamoto, T.~Kuroki, T.~Matsuo and S.~Shinohara,
\emph{Mean field approximation of IIB matrix model and emergence of
four-dimensional space-time},
\emph{Nucl.\ Phys.} {\bf B~647} (2002) 153
[{\tt hep-th/0204240}].




\bibitem{Kawai:2002ub}
  H.~Kawai, S.~Kawamoto, T.~Kuroki and S.~Shinohara,
\emph{Improved perturbation theory and four-dimensional space-time 
in IIB matrix model},
\emph{Prog.\ Theor.\ Phys.} {\bf 109} (2003) 115
[{\tt hep-th/0211272}].



\bibitem{Aoyama:2006di}
  T.~Aoyama, H.~Kawai and Y.~Shibusa,
\emph{Stability of 4-dimensional space-time 
from IIB matrix model via improved mean field approximation},
\emph{Prog.\ Theor.\ Phys.} {\bf 115} (2006) 1179
[{\tt hep-th/0602244}].


\bibitem{Aoyama:2006rk}
  T.~Aoyama and H.~Kawai,
\emph{Higher order terms of improved mean field approximation for IIB matrix
model and emergence of four-dimensional space-time},
\emph{Prog.\ Theor.\ Phys.} {\bf 116} (2006) 405
[{\tt hep-th/0603146}].




\bibitem{Aoyama:2006je}
  T.~Aoyama and Y.~Shibusa,
\emph{Improved perturbation method and its application 
to the IIB matrix model},
\emph{Nucl.\ Phys.} {\bf B~754} (2006) 48
[{\tt hep-th/0604211}].



\bibitem{Nishimura:2002va}
  J.~Nishimura, T.~Okubo and F.~Sugino,
\emph{Convergence of the Gaussian expansion method
in dimensionally reduced Yang-Mills integrals},
\emph{JHEP} {\bf 10} (2002) 043
[{\tt hep-th/0205253}].


\bibitem{Nishimura:2003gz}
  J.~Nishimura, T.~Okubo and F.~Sugino,
\emph{Testing the Gaussian expansion method 
in exactly solvable matrix models},
{\em JHEP} {\bf 10} (2003) 057
[{\tt hep-th/0309262}].



\bibitem{Aoyama:2010ry}
  T.~Aoyama, J.~Nishimura and T.~Okubo,
\emph{Spontaneous breaking of the rotational symmetry 
in dimensionally reduced super Yang-Mills models},
\emph{Prog.\ Theor.\ Phys.} {\bf 125} (2011) 537
[{\tt arXiv:1007.0883}].

\bibitem{Ambjorn:2000dx}
J.~Ambjorn, K.~N.~Anagnostopoulos, W.~Bietenholz, T.~Hotta and J.~Nishimura,
\emph{Monte Carlo studies of the IIB matrix model at large N},
\emph{JHEP} {\bf 07} (2000) 011
[\hepth{0005147}].


\bibitem{AAN} 
K.~N.~Anagnostopoulos, T.~Azuma, and J.~Nishimura,
work in progress.





\bibitem{NV}
  J.~Nishimura and G.~Vernizzi,
\emph{Spontaneous breakdown of Lorentz invariance in IIB matrix model},
\emph{JHEP} {\bf 04} (2000) 015
[{\tt hep-th/0003223}];
\emph{Brane world generated dynamically from string type IIB matrices,}
\prl{85}{2000}{4664}
[\hepth{0007022}].



\bibitem{Krauth:1998xh}
  W.~Krauth, H.~Nicolai and M.~Staudacher,
\emph{Monte Carlo approach to M theory},
\plb{431}{1998}{31}
[{\tt hep-th/9803117}].



\bibitem{Hotta:1998en}
  T.~Hotta, J.~Nishimura and A.~Tsuchiya,
\emph{Dynamical aspects of large N reduced models},
\emph{Nucl.\ Phys.} {\bf B~545} (1999) 543
[{\tt hep-th/9811220}].


\bibitem{AW}
  P.~Austing and J.~F.~Wheater,
\emph{The convergence of Yang-Mills integrals},
\emph{JHEP} {\bf 02} (2001) 028
[\hepth{0101071}];
%
\emph{JHEP} {\bf 04} (2001) 019
[\hepth{0103159}].


\bibitem{Ambjorn:2000bf}
  J.~Ambjorn, K.~N.~Anagnostopoulos, W.~Bietenholz, T.~Hotta and J.~Nishimura,
\emph{Large N dynamics of dimensionally reduced 4D SU(N) super Yang-Mills
theory},
{\em JHEP} {\bf 07} (2000) 013
[{\tt hep-th/0003208}].


\bibitem{Burda:2000mn}
  Z.~Burda, B.~Petersson and J.~Tabaczek,
\emph{Geometry of reduced supersymmetric 4D Yang-Mills integrals},
{\em Nucl. Phys.} {\bf B~602} (2001) 399
[{\tt hep-lat/0012001}].

\bibitem{Ambjorn:2001xs}
  J.~Ambjorn, K.~N.~Anagnostopoulos, W.~Bietenholz, 
F.~Hofheinz and J.~Nishimura,
\emph{On the spontaneous breakdown of Lorentz symmetry in matrix models of
superstrings},
{\em Phys. Rev.}  {\bf D~65} (2002) 086001
[{\tt hep-th/0104260}].


\bibitem{sign}
K.~N.~Anagnostopoulos and J.~Nishimura,
\emph{New approach to the complex-action problem 
and its application to a nonperturbative study of superstring theory,}
\prd{66}{2002}{106008}
[\hepth{0108041}].
%

\bibitem{Nishimura:2001sq}
  J.~Nishimura,
\emph{Exactly solvable matrix models for the dynamical generation of  
space-time in superstring theory},
{\em Phys. Rev.}  {\bf D~65} (2002) 105012
[{\tt hep-th/0108070}].





\bibitem{Nishimura:2004ts}
  J.~Nishimura, T.~Okubo and F.~Sugino,
\emph{Gaussian expansion analysis of a matrix model with the spontaneous
breakdown of rotational symmetry},
\emph{Prog.\ Theor.\ Phys.} {\bf 114} (2005) 487
[{\tt hep-th/0412194}].




\bibitem{Anagnostopoulos:2010ux}
  K.~N.~Anagnostopoulos, T.~Azuma and J.~Nishimura,
\emph{A general approach to the sign problem: The factorization method with
multiple observables},
\emph{Phys.\ Rev.} {\bf D~83} (2011) 054504 
 [{\tt arXiv:1009.4504}];\\
%
\emph{A practical solution to the sign problem in a matrix model for dynamical
compactification},
[{\tt arXiv:1108.1534}].



\bibitem{Kabat:2000hp}
D.~Kabat and G.~Lifschytz,
{\em Approximations for strongly-coupled supersymmetric quantum mechanics},
\npb{571}{2000}{419} 
[{\tt hep-th/9910001}].


\bibitem{blackholes}
D.~Kabat, G.~Lifschytz and D.A.~Lowe,
{\em Black hole thermodynamics from calculations 
in strongly-coupled gauge theory}, 
\emph{Phys.\ Rev.\ Lett.} {\bf 86} (2001) 1426
[{\tt hep-th/0007051}];\\
D.~Kabat, G.~Lifschytz and D.A.~Lowe,
{\em Black hole entropy from non-perturbative gauge theory}, 
\emph{Phys.\ Rev.} {\bf D~64} (2001) 124015 
[{\tt hep-th/0105171}];\\
N.~Iizuka, D.~Kabat, G.~Lifschytz and D.A.~Lowe,
{\em Probing black holes in non-perturbative gauge theory}, 
\emph{Phys.\ Rev.} {\bf D~65} (2002) 024012 
[{\tt hep-th/0108006}].



\bibitem{Gauss_simpleIIB}
S.~Oda and F.~Sugino,
{\em Gaussian and mean field approximations 
for reduced Yang-Mills integrals},
\emph{JHEP} {\bf 03} (2001) 026
[{\tt hep-th/0011175}];\\
F.~Sugino,
{\em Gaussian and mean field approximations for reduced 4D supersymmetric 
Yang-Mills integral},
\emph{JHEP} {\bf 07} (2001) 014 
[{\tt hep-th/0105284}].

\bibitem{Stevenson:1981vj}
P.~M. Stevenson, {\it {Optimized perturbation theory}},  
{\em Phys. Rev.}  {\bf D~23} (1981) 2916.


\bibitem{Green:1997tn}
  M.~B.~Green and M.~Gutperle,
\emph{D particle bound states and the D instanton measure},
\emph{JHEP} {\bf 01} (1998) 005
[{\tt hep-th/9711107}].




\bibitem{Moore:1998et}
  G.~W.~Moore, N.~Nekrasov and S.~Shatashvili,
\emph{D particle bound states and generalized instantons},
\emph{Commun.\ Math.\ Phys.} {\bf 209} (2000) 77
[{\tt hep-th/9803265}].



\end{thebibliography}
\end{document}